\documentstyle[12pt]{article}
\input psfig.sty
\begin{document}

{} \hfill {\bf \large IFT/29/00}

\vskip1in

\centerline{\Large {\bf Dynamics of effective gluons$^\dagger$}}

\vskip .1in
\centerline{\small {November 12, 2000}}

\vskip .3in

\centerline{Stanis{\l}aw D. G{\l}azek}

\vskip .1in
\centerline{Institute of Theoretical Physics, Warsaw University}
\centerline{ul.  Ho{\.z}a 69, 00-681 Warsaw}

\vskip.5in
\centerline{\bf Abstract}
\vskip.1in

Renormalized Hamiltonians for gluons are constructed using a
perturbative  boost-invariant renormalization group procedure for
effective particles in  light-front QCD, including terms up to third
order. The effective gluons and  their Hamiltonians depend on the
renormalization group parameter $\lambda$,  which defines the width of
momentum space form factors that appear in the  renormalized
Hamiltonian vertices. Third-order corrections to the three-gluon
vertex exhibit asymptotic freedom, but the rate of change of the
vertex with  $\lambda$ depends in a finite way on regularization of
small-$x$ singularities. This dependence is shown in some examples,
and a  class of regularizations with two distinct scales in $x$ is
found to lead to  the Hamiltonian running coupling constant whose
dependence on $\lambda$ matches the known perturbative result from
Lagrangian calculus for the dependence of gluon three-point Green's
function on the running momentum scale at large scales.  In the Fock space
basis of effective gluons with small $\lambda$, the vertex form factors 
suppress interactions with large kinetic energy changes and thus remove 
direct couplings of low energy constituents to high  energy components 
in the effective bound state dynamics.  This structure is reminiscent 
of parton and constituent models of hadrons.

\vskip1.6in

$^\dagger$ Supported by KBN Grant No. 2 P03B 016 18.

\newpage

{\bf 1. Introduction}
\vskip.1in

Current studies of hadronic structure are guided by three physical
pictures.  The first picture is based on the constituent quark model,
which serves classification of hadrons in particle data tables~\cite{PDT}.
Quantum numbers of a hadron in this model correspond to a simple
Hamiltonian with only kinetic energy of two or three quarks and
inter-quark potentials in the hadron rest frame, with no gluons.  The
second picture is provided by the parton model for hadrons in the
infinite momentum frame~\cite{Feynman}.  Modern versions of the model
introduce a slew of quarks and gluons with distribution functions in
variable $x$ - a fraction of the hadron momentum that is carried by a
parton.  About half of the hadron momentum is carried by gluons, with
mostly small values of $x$, so that many partons can share the hadron
momentum.  Binding of partons is not described by the parton model.  In
the third picture, hadrons are considered to be excitations of a
complicated ground state (vacuum) that contains condensates of quarks
and gluons.  Understanding of hadronic structure in the third way relies
on the assumed ground state properties~\cite{SVZ}.  Despite recent
progress in experimental and theoretical studies of hadronic structure,
including lattice approach~\cite{{WilsonLattice},{Lattice2000}}, the
three basic pictures are not yet unified in a single quantitative
formulation of QCD.  To connect constituent quarks and partons with QCD
degrees of freedom, one needs a relativistic description of effective
particles in quantum field theory.

This paper describes a perturbative third-order calculation of
renormalized Hamiltonians for effective gluons in the light-front Fock
space.  The effective gluons are derived in a boost-invariant
renormalization group procedure for particles~\cite{GlazekAPP12}, which
originates in the similarity approach to renormalization of
Hamiltonians~\cite{GlazekWilson12} and the notion of vertex form factors
for extended strongly interacting particles~\cite{Erkelenz}.  The
renormalization procedure provides a connection between the canonical
quantum field theory and the concepts of bound state constituents in the
rest and infinite momentum frames.  For simplicity, this paper is
limited to gluons.  Quark effects in the gluon dynamics are mentioned
only in passing.  Gluons alone are worth a discussion since their
interactions are responsible for asymptotic freedom.  This feature
requires understanding in Hamiltonians independently of quark dynamics.
Also, asymptotically free effective gluon interactions display specific
sensitivity to regularization of small-$x$ singularities.

Section 2 presents the initial regularized Hamiltonian for gluons.  The
Hamiltonian includes ultraviolet counterterms that are calculable order
by order in the procedure described in Section 3. The procedure
introduces vertex form factors in the effective gluon interactions.  The
form factor width parameter, $\lambda$, is reduced from infinity down to
the scale of hadronic masses through a solution of a differential
equation, which eliminates large momentum transfers from the bound state
eigenvalue problem for effective gluons with small $\lambda$.  The
coupling strength of the three-gluon vertex as function of $\lambda$, is
calculated in Section 4 and analyzed in Section 5. These two Sections
show how asymptotic freedom of effective gluons emerges in the
light-front Fock space Hamiltonians for QCD.  Section 6 provides a short
summary and a brief discussion of how the effective particle calculus
can be applied to electron-hadron scattering in a simplest
approximation.

\vskip.3in
{\bf 2. Initial Hamiltonian}
\vskip.1in

The canonical light-front QCD Hamiltonian requires regularization and
counterterms~\cite{Wilson6}.  To regulate the Hamiltonian, momenta $p^+
= p^0 + p^3$ and $p^\perp = (p^1, p^2)$ are parametrized using the $+$
momentum ratios $x$ and relative transverse momenta $\kappa^\perp$ that
will be described below.  Regularization is imposed through factors that
exclude large $|\kappa^\perp|$ and small $x$, preserving all kinematical
symmetries of the light-front dynamics (i.e. the Poincar\'e symmetries
of the surface $x^+=x^0 + x^3=0$ in space-time), and processes of
creation of particles from the bare vacuum are absent.  Power-counting
and renormalization strategy for the absolute co-ordinates, $x^-=x^0-x^3$
and $x^\perp= (x^1, x^2)$, or $p^+$ and $p^\perp$~\cite{Wilson6}, are
modified when one goes over to the variables $x$ and $\kappa^\perp$.
However, key features remain similar and perturbative results described
in the next Sections agree with expectation that the ultraviolet
renormalization of light-front Hamiltonians involves functions of $x$.

\vskip.3in
{\bf 2.a} Canonical terms
\vskip.1in

The classical Lagrangian density for gluon fields is

$$ {\cal L} = - {1 \over 2} tr F^{\mu \nu} F^{\mu \nu} \quad , \eqno(2.1)
$$

\noindent where $F^{\mu \nu} = \partial^\mu A^\nu - \partial^\nu A^\mu +
i g [A^\mu, A^\nu]$, and $A^\mu = A^{a \mu} t^a$ with $ [t^a,t^b] = i
f^{abc} t^c$.  The Lagrangian implies equations of motion, $\partial_\mu
F^{\mu \nu} = ig[F^{\mu \nu}, A_\mu]$, and for fields satisfying these
equations the canonical energy-momentum density tensor is $T^{\mu \nu} =
-F^{a \mu \alpha} \partial^\nu A^a_\alpha + g^{\mu \nu} F^{a \alpha
\beta} F^a_{\alpha \beta}/4$.

In the gauge $A^{a +} = 0$, the Lagrange equations constrain $A^-$ to
$\tilde A^- = (1/\partial^+)2\partial^\perp A^\perp - (2/ \partial^{+ \,
2})ig [ \partial^+ A^\perp, A^\perp]$ and the independent field degrees
of freedom are $A^\perp$.  The first term in $\tilde A^-$ is independent
of the coupling constant $g$, and can by definition be included in a new
constrained field $A^\mu = [ A^+=0, A^- = (1/ \partial^+)2\partial^\perp
A^\perp, A^\perp ] $. The second term can be kept explicitly as part of
interactions.  Using this convention and freely integrating by parts,
one obtains an expression for light-front energy of the constrained
gluon field

$$ P^- = {1 \over 2}\int dx^- d^2 x^\perp {\cal H}|_{x^+=0} \quad ,
\eqno(2.2) $$

\noindent where ${\cal H} = T^{+ -}$ and

$$ T^{+ -} = {\cal H}_{A^2} + {\cal H}_{A^3} + {\cal H}_{A^4} + {\cal
H}_{[\partial A A]^2} \quad, \eqno(2.3) $$

\noindent with

$$ {\cal H}_{A^2} = - {1\over 2} A^\perp (\partial^\perp)^2 A^\perp
\quad , \eqno(2.3a)$$

$$ {\cal H}_{A^3} = g \, i\partial_\alpha A_\beta^a [A^\alpha,A^\beta]^a
\quad , \eqno(2.3b)$$

$$ {\cal H}_{A^4}= - {1\over 4} g^2 \,
[A_\alpha,A_\beta]^a[A^\alpha,A^\beta]^a \quad ,\eqno(2.3c)$$

$$ {\cal H}_{[\partial A A]^2} = {1\over 2}g^2 \,
[i\partial^+A^\perp,A^\perp]^a {1 \over (i\partial^+)^2 }
[i\partial^+A^\perp,A^\perp]^a \quad . \eqno(2.3d)$$

\noindent This expression is a candidate for further consideration in
analogy to QED~\cite{{BKS},{Yan},{BRS},{LB}}.  A heuristic expression
for the quantum gluon energy operator is obtained by substitution

$$ A^\mu = \sum_{\sigma c} \int [k] \left[ t^c \varepsilon^\mu_{k\sigma}
a_{k\sigma c} e^{-ikx} + t^c \varepsilon^{\mu *}_{k\sigma}
a^\dagger_{k\sigma c} e^{ikx}\right]_{x^+=0} \quad , \eqno(2.4) $$

\noindent where $ [k] = \theta(k^+) k^+ d^2 k^\perp/(16\pi^3 k^+) $ and
$\varepsilon^\mu_{k\sigma} = (\varepsilon^+_{k\sigma}=0,
\varepsilon^-_{k\sigma}= 2k^\perp \varepsilon^\perp_\sigma/k^+,
\varepsilon^\perp_\sigma) $. $\sigma$ numbers gluon spin polarization
states and $c$ is a color index.  The creation and annihilation
operators satisfy commutation relations

$$ \left[ a_{k\sigma c}, a^\dagger_{k'\sigma' c'} \right] = k^+
\not\!\delta(k - k') \,\, \delta^{\sigma \sigma'} \, \delta^{c c'} \quad
, \eqno(2.4a) $$

\noindent where $\not\!\delta(p) = 16 \pi^3 \delta(p^+) \delta(p^1)
\delta(p^2)$, and commutators among all $a$s, and among all
$a^\dagger$s, vanish.  For all momenta, spins and colors, $ a_{k\sigma
c} |0\rangle \equiv 0$ and $a^\dagger_{k\sigma c}$ creates bare gluons
from the state $|0\rangle$.

The plain insertion of Eq.  (2.4) into $P^-$ produces terms with
creation and annihilation operators appearing in all possible orders.
All terms are then ordered so that creation operators stand to the left
of annihilation operators.  The process of ordering produces commutators
of creation and annihilation operators, which lead to diverging
integrals.  All such terms can be dropped at this stage entirely since
they will be either removed by regularization, in the case of modes with
$k^+=0$ or, after regularization, they will be replaced by well defined
mass counterterms that result from a renormalization group procedure and
contain free additive finite parts.

The ordered operator, denoted $P^-_{quantum}$, is highly divergent.  For
example, a correction order $g^2$ to the free energy $k^-=k^{\perp \,
2}/k^+$ of a single bare gluon state $|k\sigma c\rangle =
a^\dagger_{k\sigma c}|0\rangle$, diverges due to integration over an
infinite range of transverse momenta of virtual gluons that appear in
the intermediate states of second order perturbation theory.  The energy
correction diverges also due to small $x$ singularities.  Namely, the
gluon momentum $k^+$ can be shared by two intermediate gluons carrying
fractions $x$ and $1-x$.  The sum over intermediate states involves an
integral over $x$ from 0 to 1 while the polarization vectors of
intermediate gluons provide $x$ and $1-x$ in denominator of the
integrand, cf.  $\varepsilon^\mu_{k\sigma}$ in Eq.  (2.4).  As another
example, the product $P^-_{quantum} P^-_{quantum} $ is even more
divergent than the energy correction because it does not contain the
energy denominator that reduces the contribution of intermediate states
with large momenta in perturbation theory.  Consequently, $\exp {( -i
P^-_{quantum} x^+/2)} $ as a candidate for a unitary evolution operator
in time $x^+$, is not defined before one regulates $P^-_{quantum}$ by
limiting the range of momentum that the bare gluons may have.

\vskip.3in
{\bf 2.b} Regularization
\vskip.1in

The first step in regularization procedure is made by limiting the range
of momentum integration in Eq.  (2.4), cf.  Ref.~\cite{Wilson6}.
Let $|k^\perp| < \Omega$ and $k^+ > \epsilon^+$, with understanding that
$\Omega \rightarrow \infty$ and $\epsilon^+ \rightarrow 0$ when the
regularization is being removed.

The lower bound on $k^+$ implies that the regulated expression for
$P^-_{quantum}$, denoted by $P^-_{\Omega \epsilon^+}$, does not contain
any terms with exclusively creation, or annihilation operators.  Such
terms would be forced by a translationally invariant integral over $x^-$
to preserve momentum $k^+$, while the momentum they would have to
create, or destroy, is at least $n \epsilon^+$, where $n$ denotes the
number of creation, or annihilation operators in such terms,
respectively.  These two conditions are incompatible.  Hence,
$P^-_{\Omega \epsilon^+}$ does not contain terms that could alter the
bare vacuum state $|0\rangle$.  The coupling constant $g$ is assumed to
be sufficiently small for stability of the regularized theory built on
top of $|0\rangle$, cf.~\cite{Susskind}.

The limits on absolute momenta in $P^-_{\Omega \epsilon^+}$ violate
boost invariance of light-front dynamics, and in the term (2.3d) with
inverse powers of $\partial^+$, one may still have 0 in denominator.  To
eliminate the violation of boost invariance and regulate the
$1/\partial^+$ singularities, $P^-_{\Omega \epsilon^+}$ is further
curbed through the following step.

Interaction terms in $P^-_{\Omega \epsilon^+}$ are modified so that {\it
changes} of the particle momenta are limited.  In this work, the
transverse momentum changes are limited by a parameter $\Delta$, and
changes of $x$ by a parameter $\delta$. As an example, it is useful to
consider the three-gluon vertex,

$$ H_{A^3 \, \Omega \epsilon^+ } = \sum_{123}\int[123]
\not\!\delta(k_1+k_2-k_3)\, \left[C_{\Omega \epsilon^+} (123)\,
a^\dagger_1 a^\dagger_2 a_3 + C_ {\Omega \epsilon^+} ^* (123)\,
a^\dagger_3 a_2 a_1 \right] \quad , \eqno(2.5)$$

\noindent Momentum conservation implies that one can write $k_1^+ = x_1
k_3^+$, $k_1^\perp = x_1 k_3^\perp +\kappa_{12}^\perp$, $k_2^+ = x_2
k_3^+$, $k_2^\perp = x_2 k_3^\perp -\kappa_{12}^\perp$, with $x_1 + x_2
= 1$, and $\kappa_{12}^\perp = x_2 k_1^\perp - x_1
k_2^\perp$.  $x_1$, $x_2$, and $\kappa^ \perp_{12}$ are invariant under
seven kinematical transformations of light-front dynamics.  It is
helpful to call the momentum carried together by all annihilated or
created particles in a single vertex a {\it parent} momentum in the
vertex.  In the vertex (2.5), $k_3$ is a parent momentum.  Also, a slash
in a subscript is used below to indicate that the momentum before the
slash sign is considered to be a daughter of the parent momentum after
the slash sign.  For example, $k^\perp_1 = x_{1/3} k^\perp_3 + 
\kappa^\perp_{1/3}$, where $x_{1/3} = x_1/x_3$ and
$\kappa^\perp_{1/3} = \kappa^\perp_{12}$.

The momentum changes in the vertex (2.5), are limited by inserting a
factor $r_{\Delta \delta}(\kappa_{i/3}^{\perp \, 2}, x_{i/3})$ for each
creation and annihilation operator.  Since $k_3$ is a parent
momentum, $x_{3/3} = 1$, and $\kappa^\perp_{3/3} = 0$, and the
regularization factor for the term (2.5) equals $r_{\Delta
\delta}(\kappa^{\perp \, 2},x) r_{\Delta \delta}(\kappa^{\perp \, 2},
1-x)$, where $x=x_{1/3}$ and $\kappa^\perp = \kappa^\perp_{12}$.
Factors $r_{\Delta \delta}$ are chosen to have the form

$$ r_{\Delta \delta}(\kappa^{\perp \, 2}, x) = r_\Delta( \kappa^{\perp
\, 2}) r_\delta(x) \theta(x) \quad , \eqno(2.6) $$

\noindent where

$$ r_\Delta(z) = \exp{(- z/\Delta^2)} \quad , \eqno(2.7) $$

\noindent and $r_\delta(x)$ suppresses the region of $x$ smaller than
$\delta$.  That $r_\Delta(z)$ falls off exponentially is a guarantee for
ultraviolet convergence of all transverse momentum integrals that appear
in perturbation theory.  Integrals that behave as $\ln{\Delta}$ or
$\Delta^n$ with positive $n$ for $\Delta \rightarrow \infty$, will be
called ultraviolet divergent.  The small-$x$ regulating function,
$r_\delta(x)$, must vanish sufficiently quickly for $x \rightarrow 0$ to
regulate all small-$x$ singularities.  In a sense to become clear in
Section 5, the factors $r_\delta$ considered in this work lie in the
vicinity of two cases,

$$ r_\delta(x) = x/(x+\delta) \quad , \eqno(2.8) $$

\noindent and

$$ r_\delta(x) = \theta(x-\delta) \quad . \eqno(2.9) $$

\noindent Integrals that behave as $\ln{\delta}$ or $\delta^n$ with
negative $n$, will be called small-$x$ divergent.  Mixing of the
ultraviolet and small-$x$ regularizations through expressions of the
type $\exp{[-(\kappa^{\perp \, 2}/x)/\Delta^2]}$, is discussed in
Appendix E.

Every creation and annihilation operator in all vertices in the operator
$P^-_{\Omega \epsilon^+}$ is supplied with a factor $r_{\Delta\delta}$,
$k_3$ being replaced by the corresponding parent momentum.  Terms that
contain four operators are recast as contracted products of terms with
only three operators, which are already regulated.  Details of the
quartic terms are given in Appendix A. The instantaneous terms
containing inverse powers of $\partial^+$ are regulated in the same way,
by interpreting the momentum that is transfered along the inverted
$\partial^+$ as a momentum carried by a virtual particle that connects
two vertices~\cite{{Yan},{BRS}}.  The fully regulated operator
$P^-_{\Omega \epsilon^+}$ is denoted by $\left[P^-_{\Omega
\epsilon^+}\right]_{\Delta\delta}$.

In the last step of defining the initial Hamiltonian,
$H_{\Delta\delta}$, the limits of $\Omega \rightarrow \infty$ and
$\epsilon^+ \rightarrow 0$ are taken with $\Delta$ and $\delta$ kept
constant,

$$ H_{\Delta \delta} = \lim_{\Omega \rightarrow \infty} \lim_{\epsilon^+
\rightarrow 0} \left[P^-_{\Omega \epsilon^+}\right]_{\Delta\delta} \quad
+ \quad X_{\Delta \delta} \quad . \eqno(2.10) $$

\noindent $X_{\Delta \delta}$ denotes counterterms, which need to be
found.  The renormalization group procedure that provides means for
finding them order by order in perturbation theory, is described in next
Sections.  The initial Hamiltonian has then the form, cf.  Eq.  (2.3),

$$ H_{\Delta \delta} = H_{A^2} + H_{A^3} + H_{A^4} + H_{[\partial A
A]^2} + X_{\Delta \delta} \quad , \eqno(2.11) $$

\noindent where, for example,

$$ H_{A^2} = \sum_{\sigma c} \int [k] {k^{\perp \, 2} \over k^+}
a^\dagger_{k\sigma c}a_{k\sigma c} , \eqno(2.12a) $$

\noindent and

$$ H_{A^3} = \sum_{123}\int[123] \not\!\delta(p^\dagger - p)\,\tilde
r_{\Delta\delta}(3,1) \left[g\,Y_{123}\, a^\dagger_1 a^\dagger_2 a_3 +
g\,Y_{123}^*\, a^\dagger_3 a_2 a_1 \right] . \eqno(2.12b) $$

\noindent These two terms are quoted from the Appendix A, where all
regulated canonical terms are listed and the notation is explained.
$X_{\Delta \delta}$ will be discussed below and in next Sections.  Note
that the free Hamiltonian (2.12a) contains no regularization.  This is
necessary to preserve kinematical light-front symmetries. Differences 
between the bare three-gluon vertex (2.12b) and effective gluon vertex
from Section 5, are described there.

\vskip.3in
{\bf 3. Renormalization group procedure for particles}
\vskip.1in

In Eq.  (2.11), $H_{\Delta \delta}$ is expressed in terms of the
creation and annihilation operators for bare gluons in regulated local
theory.  This Section describes the renormalization group
procedure~\cite{GlazekAPP12} that is used in Section 4 to re-write the
initial Hamiltonian $H_{\Delta \delta}$ in terms of operators that
create or annihilate effective gluons, instead of bare ones.  The
effective gluon operators are obtained by applying a unitary
transformation ${\cal U}_\lambda$ to the initial bare operators.  The
effective operators depend on the parameter $\lambda$ that labels $\cal
U_\lambda$.

The parameter $\lambda$ has dimension of mass and distinguishes
different kinds of effective gluons according to the following rule.
{\it Effective gluons of type $\lambda$ can change their relative
motion kinetic energy through a single effective interaction by no 
more than about $\lambda$.} The transformation $\cal U_\lambda$ is 
mathematically designed in perturbation theory so that resulting 
interaction terms contain vertex form factors and the latter limit
the kinetic energy changes by their width parameter, which equals 
$\lambda$.  All Hamiltonians with different $\lambda$s are equal and 
the re-writing does not introduce any change in the theory, although 
the same Hamiltonian appears differently when expressed in terms of 
different gluons.  For brevity, the effective gluons corresponding 
to some value of $\lambda$ are referred to as gluons of width $\lambda$.

\vskip.3in
{\bf 3.a} General features
\vskip.1in

The form factor width parameter $\lambda$ greatly differs from the
regularization cutoffs, because it may be kept finite, even small, while
the cutoffs have to be made extremely large to approximate the initial
theory.  Even if the expansion in terms of bare particles is hopelessly
complicated, a hadron may still have a well defined, convergent
expansion in the basis of effective constituents with small width
$\lambda$.

Thanks to the vertex form factors, in the Fock space basis built from
gluons of width $\lambda$, the effective Hamiltonian matrix elements
quickly tend to zero when the effective gluons change kinetic energy across
the matrix element by more than $\lambda$.  Therefore, the effective
Hamiltonian matrix is narrow.  This is important for application to
bound state physics because eigenstates of narrow matrices may have a
small number of dominant components~\cite{GWW}.  In the case of hadrons,
the constituent model suggests that the intricate complexity of QCD is
buried in the structure of constituents and their interactions, while
the number of effective constituents is small.  The success of
perturbative QCD in reproducing changes of deep inelastic structure
functions with momentum transfer down to fairly small values, suggests
that the structure of effective constituents can be approximated using
perturbation theory.  The idea of effective particles is by no means
new~\cite{Melosh}.  New element is the renormalization group procedure
for effective particles in QCD.

The renormalization group procedure provides means to find counterterms
$X_{\Delta \delta}$ that have to be included in the initial Hamiltonian
to compensate for the spurious effects of the ultraviolet
regularization.  One takes advantage of the form factors in the
Hamiltonian vertices in analogy to Ref.~\cite{GlazekWilson12}.  The
narrow dynamics can smear states of effective particles with finite
energy, only by less than $n\lambda$ towards high energies in $n$th
order perturbation theory.  For to raise the free energy of a state by
$n \lambda$, the effective interaction must act on the state about $n$
times when the form factors die out exponentially for energy changing
by more than $\lambda$.  The highest order $n$ that is still independent
of $\Delta$, approaches $\infty$ when $\Delta \rightarrow \infty$.
Therefore, to obtain the ultraviolet regularization independent results,
at least in perturbation theory to all orders, it is sufficient to
demand that the Hamiltonian coefficients in front of creation and
annihilation operators with finite $\lambda$ are independent of the
ultraviolet regularization.  Hence, one can read the structure of
ultraviolet counterterms from the coefficients, see Eq.  (3.11).
However, the effects of small-$x$ regularization are not under control
of the renormalization group procedure and the coefficients with finite
$\lambda$s may depend on $r_\delta(x)$.  The dependence on $\delta
\rightarrow 0$ sometimes drops out but finite effects may remain, as
will be shown on examples in Section 5.

In the perturbative renormalization group procedure for deriving
effective particles and their interactions, one never encounters genuine
infrared singularities associated with small energy denominators.  This
is explained below Eq.  (3.10), where the differences of invariant
masses are taken care of in analogy to differences of energies in the
similarity renormalization group procedure for
Hamiltonians~\cite{GlazekWilson12}.  The perturbative denominators are
effectively limited from below by $\lambda$.  The non-perturbative part
of dynamics with relative motion kinetic energy changes smaller than 
$\lambda$, is first tackled when one proceeds to solve the effective 
Schr\"odinger equation. Since the form factors keep the effective dynamics 
in a well defined range of energies, numerical methods may apply in 
finding approximate solutions to the full theory~\cite{GWW}.

\vskip.3in
{\bf 3.b} Construction of ${\cal H}_\lambda$~\cite{GlazekAPP12}
\vskip.1in

Let $a$ commonly denote the bare operators, $a_{k\sigma c}$ or
$a^\dagger_{k\sigma c}$.  Operators $a$ are transformed by the unitary
operator $\cal U_\lambda$ into operators $a_\lambda$ that create or
annihilate effective particles of width $\lambda$, with identical
quantum numbers.

$$ a_\lambda \,\, = \,\, {\cal U}_\lambda \,\, a \,\, {\cal
U}^\dagger_\lambda \,\, . \eqno(3.1) $$

\noindent The initial Hamiltonian $H_{\Delta \delta}$ is re-written in
terms of $a_\lambda$, $H_{\Delta \delta} = H_\lambda(a_\lambda)$.  If
quarks were included, $H_{\Delta \delta}$ would correspond to the QCD
Hamiltonian written in terms of canonical quarks and gluons, associated
with bare partons, or bare currents.  $H_\lambda$ for $\lambda$
comparable with masses would represent the same Hamiltonian written in
terms of constituent quarks and gluons.  Applying ${\cal U}_\lambda$,
one obtains

$$ {\cal H}_\lambda \equiv H_\lambda(a) = {\cal U}^\dagger_\lambda
H_{\Delta \delta} {\cal U}_\lambda \quad . \eqno(3.2) $$

\noindent ${\cal H}_\lambda$ has the same coefficient functions in front
of products of $a$s as the effective $H_\lambda$ has in front of the
unitarily equivalent products of $a_\lambda$s.  Differentiating ${\cal
H}_\lambda$ with respect to $\lambda$, one obtains

$$ {\cal H}'_\lambda = - [{\cal T}_\lambda, \, {\cal H}_\lambda] \quad ,
\eqno(3.3) $$

\noindent where ${\cal T}_\lambda = {\cal U}^\dagger_\lambda \,{\cal
U}'_\lambda$.  ${\cal T}_\lambda$ is constructed below using the notion
of vertex form factors.  For example, if an operator without form
factors has the structure

$$ \hat O_{\lambda} = \int [k_1 k_2 k_3] \,\, V_\lambda(1,2,3) \,\,
a^\dagger_{\lambda k_1} a^\dagger_{\lambda k_2} a_{\lambda k_3} \,\, ,
\eqno(3.4) $$

\noindent the operator with form factors is written as $f_\lambda {\hat
O}_{\lambda}$ and has the structure

$$ f_\lambda {\hat O}_{\lambda} = \int [k_1 k_2 k_3] \,\,
f_\lambda({\cal M}_{12},{\cal M}_3)\,\, V_\lambda(1,2,3) \,\,
a^\dagger_{\lambda k_1} a^\dagger_{\lambda k_2} a_{\lambda k_3} \,\,,
\eqno(3.5) $$

\noindent where

$$ f_\lambda({\cal M}_{12},{\cal M}_3) = \exp[-({\cal M}_{12}^2 - {\cal
M}_3^2)^2/\lambda^4)] \,\, . \eqno(3.6)$$

\noindent For any operator $\hat O$ expressible as a linear combination
of products of creation and annihilation operators, $f \hat O$ contains
a form factor $f_\lambda({\cal M}_c, {\cal M}_a)$ in front of each
product, where ${\cal M}_c$ and ${\cal M}_a$ stand for the total free
invariant masses of particles created (c) and annihilated (a) through
the product, respectively.  For gluons in Eq.  (3.6), ${\cal M}_3 =0$.

The relative motion kinetic energy changes in interaction vertices of 
effective particles,
are limited by demanding that $H_\lambda = f_\lambda G_\lambda$, with
some unknown $G_\lambda$.  Then, one derives equations for $G_\lambda$
that result from the choice for $f_\lambda$ and some definition of
${\cal T}_\lambda$.  In practice, one first uses ${\cal U}_\lambda$ to
transform the Hamiltonian to ${\cal H}_\lambda = f_\lambda {\cal
G}_\lambda$ and then one calculates coefficients of $a$s in ${\cal
G}_\lambda$, which are the same as the coefficients of $a_\lambda$ in
$G_\lambda$.  This calculation includes the construction of ${\cal
T}_\lambda$, and proceeds as follows (the subscript $\lambda$ is dropped
for simplicity of notation).

$$ {\cal H}' = f' {\cal G} + f {\cal G}' = - [{\cal T}, \, {\cal G}_0] -
[{\cal T}, \, {\cal G}_I] \,\, . \eqno(3.7) $$

\noindent ${\cal G}$ is split into two parts.  The free part ${\cal
G}_0$ is bilinear in $a$s and independent of the coupling constant $g$.
The remaining, interaction dependent part is denoted by ${\cal G}_I$.
In the present work, ${\cal G}_0$ is taken to be independent of
$\lambda$.  Eq.  (3.7) contains two unknowns, ${\cal T}$ and ${\cal
G}_I$.  Without loss of generality, one assumes that ${\cal T}
\rightarrow 0$ when ${\cal G}_I \rightarrow 0$, and one expands
operators in powers of ${\cal G}_I$ with the goal of enabling the
procedure to work order by order.  The expansion of ${\cal T}$ starts
from the term of order ${\cal G}_I$.  Changes of ${\cal G}_I$ with
$\lambda$ should start from second power.  If

$$ f {\cal G}'_I = - f [{\cal T}, \, {\cal G}_I] \,\, , \eqno(3.8) $$

\noindent then

$$ [{\cal T}, \, {\cal G}_0] = [(1 - f) {\cal G}_I]' \,\, , \eqno(3.9)
$$

\noindent and

$$ {\cal G}'_{I} \,\, = \,\, \left[ f{\cal G}_I, \, \left\{ (1-f){\cal
G}_I \right\}'_{{\cal G}_0} \right] \,\, , \eqno(3.10) $$

\noindent where the curly bracket with subscript ${\cal G}_0$,
indicates the solution for ${\cal T}$ that follows from Eq.  (3.9).
The choice of $f$ made above implies that perturbation theory for
${\cal T}$ and $ {\cal G}$ does not lead to small energy differences
in denominators, since $1-f$ vanishes quadratically with the energy
difference.  ${\cal G}_{I}$ contains only connected interactions
because Eq.  (3.10) has a commutator on the right-hand side.  The
initial condition for Eq.  (3.10) is provided by $H_{\Delta \delta}$,
so that Eq.  (3.10) in the integral form reads

$$ {\cal G}_\lambda \,\, = \,\, H_{\Delta \delta} \,\, + \,\,
\int_\infty^\lambda ds \,\left[ f_s{\cal G}_{Is}, \, \left\{(1-f_s){\cal
G}_{Is} \right\} _{{\cal G}_0} \right] \,\, , \eqno(3.11) $$

\noindent which allows one to find the counterterms $X_{\Delta \delta}$
using the condition that they remove dependence on regularization from
the second term for finite $\lambda$ and relative momenta of interacting 
particles.  Finally, ${\cal H}_\lambda = f_\lambda {\cal G}_\lambda$.

\vskip.3in
{\bf 3.c} Perturbation theory for ${\cal G}_{I \lambda}$
\vskip.1in

This Section contains formulas used in Section 4 for solving Eq.  (3.11)
in perturbation theory up to third order.  Since the perturbative
expansion has formally the same structure as in scalar
theory~\cite{phi3}, only main steps are listed.  In the first step,
${\cal G}_{I \lambda}$ is expanded into a series of terms $\tau_n \sim
g^n$,

$$ {\cal G}_I \quad = \quad \sum_{n=1}^\infty \, \tau_n \quad .
\eqno(3.12) $$

It immediately follows from Eq.  (3.11) that $\tau_1$ is independent of
$\lambda$ and equal to the second term in the initial Hamiltonian from
Eq.  (2.11), i.e.  (2.12b).  One has $ \tau_1 = \alpha_{21} +
\alpha_{12} $, where $\alpha_{21}$ denotes the first and $\alpha_{12}$
the second term on the right-hand side of Eq.  (2.12b).  The left
subscript denotes the number of creation and the right subscript the
number of annihilation operators.  The corresponding Hamiltonian
interaction term is obtained by multiplying the integrand in Eq.
(2.12b) by $f_\lambda = \exp{[ - (k_1+k_2)^4/\lambda^4 ]}$ and
transforming $a$s into $a_\lambda$s.

For $\tau_2=\beta_{11} + \beta_{31} + \beta_{13} + \beta_{22}$, Eq.
(3.10) implies

$$ \tau_2' = [\{f'\tau_1\}, f\tau_1] \equiv f_2 \, [\tau_1 \tau_1] \quad
, \eqno(3.13) $$

\noindent where $f_2 = \{f'\}f - f\{f'\}$, with understanding that the
first factor $f$ in all terms of $f_2$ is for the first $\tau$ in the
square bracket, and the second factor $f$ in all terms of $f_2$ is for
the second $\tau$ in the bracket.  The square bracket denotes all
connected terms that result from contractions replacing products $a_i
a_j^\dagger$ by commutators $[a_i, a_j^\dagger]$.  The solution for
$\tau_2$ is then

$$ \tau_{2 \lambda} = {\cal F}_{2 \lambda} [\tau_1 \tau_1] + \tau_{2
\infty} \quad, \eqno(3.14) $$

\noindent where ${\cal F}_{2\lambda} = \int_\infty^\lambda f_2$ depends
on incoming and outgoing momenta in the two vertices formed by the
operators in the square bracket.  In the sequence $a\tau_{ab} b
\tau_{bc} c$, the three successive configurations of particle momenta
are labeled by $a$, $b$ and $c$.  To write down compact expressions for
${\cal F}_{2\lambda}$, the symbol $uv = {\cal M}^2_{uv} - {\cal
M}^2_{vu}$ is defined, where ${\cal M}^2_{uv}$ denotes the free
invariant mass of a set of particles from configuration $u$ that are
connected to the particles in configuration $v$ by an interaction
$\tau_{uv}$ in the sequence $u \tau_{uv} v$.  Spectators of the
interaction $\tau_{uv}$ do not count.  In this notation,

$$ f_\lambda({\cal M}_{ab},{\cal M}_{ba}) = \exp{[-(ab^2/\lambda^4)]}
\equiv f_{ab} \quad . \eqno(3.15) $$

\noindent The parent momentum for the vertex connecting two
configurations $u$ and $v$, is denoted by $P_{uv}$, and in the following
equations $p_{uv}$ is written in place of $P^+_{uv}$.  In all
expressions, the minus component of momentum of every gluon is given by
the eigenvalue of ${\cal G}_0 = H_{A^2}$ from Eq.  (2.12a), i.e.  $k^- =
k^{\perp \, 2}/k^+$.  Thus, with the Gaussian vertex form factors,

$$ {\cal F}_2 (a, b, c) = { p_{ba} ba + p_{bc} bc \over ba^2 + bc^2 } \,
\left[ f_{ab} f_{bc} - 1 \right] \quad . \eqno(3.16) $$

\noindent In Eq.  (3.14),

$$ \tau_{2 \infty} = H_{A^4} + H_{[\partial A A]^2} + X_{\Delta \delta
2} \quad , \eqno(3.17) $$

\noindent where $X_{\Delta \delta 2}$ denotes all ultraviolet
counterterms proportional to $g^2$.

For third order terms, $ \tau_3 = \gamma_{21} + \gamma_{12} +
\gamma_{41} + \gamma_{14} + \gamma_{32} + \gamma_{23}$, Eq.  (3.10)
gives

$$ \tau_3' = [f\tau_1, \, \{(1-f)\tau_2 \}'] \, + \, [\{f'\tau_1\}, \,
f\tau_2 ] \quad . \eqno(3.18) $$

\noindent After integration,

$$ \tau_{3 \lambda} = {\cal F}_{31\lambda}[ \tau_1 [\tau_1 \tau_1 ]] -
{\cal F}_{32\lambda} [[ \tau_1 \tau_1 ] \tau_1 ] + {\cal
F}_{2\lambda}[\tau_{2\infty} \tau_1 + \tau_1 \tau_{2\infty}] +
\tau_{3\infty} \, , \eqno(3.19) $$

\noindent where, for any sequence $a \tau_{ab} b \tau_{bc} c \tau_{cd} d$,

$$ {\cal F}_{32}(a,b,c,d) = - {\cal F}_{31} (d,c,b,a) \quad ,
\eqno(3.20) $$

\begin{eqnarray*}
&&{\cal F}_{31}(a,b,c,d) = {p_{cb}cb + p_{cd}cd \over cb^2 + cd^2}
\left[(p_{bd}bd + p_{ba}ba)\left[
{f_{ab}f_{bc}f_{cd}f_{bd} - 1 \over ab^2 + bc^2 + cd^2 + bd^2} -
{f_{ab}f_{bd} - 1 \over ab^2 + bd^2}\right] + \right. \\
&& \\
&&\ + \left. p_{bd}  \, {bc^2 + cd^2 \over db}
\left[ {f_{ab}f_{bc}f_{cd} - 1 \over ab^2 + bc^2 + cd^2} -
{f_{ab}f_{bc}f_{cd}f_{bd} - 1 \over ab^2 + bc^2 + cd^2 + bd^2}\right]
\right] \, .
\end{eqnarray*}
$$ \eqno(3.21)$$

\noindent The last term in Eq.  (3.19) denotes counterterms proportional
to $g^3$, i.e.  $\tau_{3\infty} = X_{\Delta \delta 3}$.  The next
Section describes calculation of $\gamma_{21}$.  The calculation
requires knowledge of $\beta_{11}$, $\beta_{22}$ and $\beta_{31}$.  For
all terms, $\pi_{ij} = \pi^\dagger_{ji}$.

\vskip.3in
{\bf 4. Interactions of effective gluons}
\vskip.1in

This Section describes derivation of effective gluon dynamics in third
order perturbation theory in the coupling constant $g_\lambda$ that
measures the strength of effective three-gluon vertex in the Hamiltonian
$H_\lambda(a_\lambda)$.  Since $g_\lambda = g + o(g^3)$, Hamiltonian
terms of order $g_\lambda$ and $g_\lambda^2$ can be calculated using
expansion in powers of $g$ before one proceeds to the third order terms
that define $g_\lambda$.  All calculations are carried out in the
framework described in Section 3.

In the re-written Hamiltonian $H_\lambda(a_\lambda)$, the terms that
have coefficients of order $1$, are

$$ H_{(0)} = \sum_{\sigma c} \int [k] {k^{\perp \, 2} \over k^+}
a^\dagger_{\lambda k\sigma c}a_{\lambda k\sigma c} \quad.  \eqno(4.1) $$

\noindent The subscript $\lambda$ indicates that $a_{\lambda k\sigma
c}$ annihilate and $a^\dagger_{\lambda k\sigma c}$ create effective
gluons of width $\lambda$. The effective gluons can also be
interpreted as having a spatial transverse width on the order of
$1/\lambda$ for moderate values of $x$. This interpretation is
explained below Eq. (4.3).

The terms in $H_\lambda(a_\lambda)$ that have coefficients order $g$,
are (see Appendix A for details of the notation)

$$ H_{(1)} = \sum_{123}\int[123] \not\!\delta(p^\dagger - p) \,
f_\lambda({\cal M}_{12},0) \, {\tilde r}_{\delta}(x_1)
\left[g\,Y_{123}\, a^\dagger_{\lambda 1} a^\dagger_{\lambda 2}
a_{\lambda 3} + g\,Y_{123}^*\, a^\dagger_{\lambda 3} a_{\lambda 2}
a_{\lambda 1} \right] , \eqno(4.2) $$

\noindent where

$$ {\tilde r}_\delta (x) = r_\delta(x) r_\delta(1-x) \quad ,
\eqno(4.3) $$

\noindent and the form factor $f_\lambda({\cal M}_{12},0) = \exp{[ -
\kappa^{\perp \, 4}_{12}/(x_1 x_2 \lambda^2)^2]}$ falls off as
function of the relative transverse momentum at a rate that depends on
$x$ carried by gluons.  For moderate values around $1/2$, the
transverse momentum width is on the order of $\lambda/2$, but for $x$
approaching 0, the transverse momentum width of the vertex becomes
very small, leading to a spread of the interaction strength in the
transverse spatial directions. Thus, the coupling of effective gluons
to the wee region is quite different from the canonical coupling in
Eq.  (2.12a).

Terms with coefficients order $g^2$ are derived by changing $a$ to
$a_\lambda$ in $ \tau_{\lambda 2} = \beta_{\lambda 11} +
\beta_{\lambda 31} + \beta_{ \lambda 13} + \beta_{\lambda 22} $, and
by inserting form factors as described in Section 3.b.  The
contribution of $\tau_{2 \infty}$ includes a counterterm induced by
the ultraviolet regularization $r_\Delta$.  Namely, to evaluate terms
$\beta_{\lambda 11}$ one needs to know the counterterm $\beta_{\infty
11}$.  It follows from Eq.  (3.14) that

$$ \beta_{\lambda 11} = \sum_{\sigma c}\int [k] \, {\mu^2_\lambda
\over k^+} a^\dagger_{k \sigma c} a_{k \sigma c} \quad , \eqno(4.4) $$

\noindent where

$$ \mu^2_\lambda = { g^2 \over 16 \pi^3 } \int_0^1 {dx \over x(1-x) }
\int d^2 \kappa^\perp \, {1\over k^+}{\cal F}_{2\lambda} (k,K,k) \,\,
2 \sum_{12} |Y_{12k}|^2 \,\, {\tilde r}^2_{k,1} \quad + \quad
\mu^2_\infty \quad , \eqno(4.5) $$

\noindent and the last term, $\mu^2_\infty$, is contributed by the
counterterm $\beta_{\infty 11}$.  The structure of $\beta_{\infty 11}$
is known here from hindsight, i.e. the regularization dependence of
the integral in Eq.  (4.5) results in a number that depends on the
function $r_\Delta(\kappa^2)$ but does not depend on the gluon quantum
numbers.  One has $ k^- = k^{\perp \, 2}/k^+$, $K^- = ({\cal M}^2 +
k^{\perp \, 2})/k^+$, ${\cal M}^2 = \kappa^{\perp \, 2}/[x(1-x)]$ and
${\tilde r}_{k,1}$ is given at the end of Appendix A. The sum over
quantum numbers of intermediate two-gluon states is $\sum_{12}
|Y_{12k}|^2 = N_c \kappa^2 [1 + 1/x^2 + 1/(1-x)^2] = \kappa^2
P(x)/[2x(1-x)]$, where $P(x)$ is the Altarelli-Parisi gluon splitting
function $P_{GG}(x)$~\cite{AP}.  $N_c = 3$ denotes the number of
colors.

Assuming that for some $\lambda = \lambda_0$ the effective gluon mass
squared should have the value $\mu^2_{0\delta}$, one can calculate the
counterterm mass $\mu^2_\infty$ from Eq.  (4.5).  The resulting
effective mass takes the form

$$ \mu^2_\lambda = \mu^2_{0\delta} + {g^2 \over 16 \pi^3 } \int_0^1
{dx \, r_{\delta \mu}(x) \over x(1-x) } \int d^2 \kappa^\perp \,
{1\over k^+}[{\cal F}_{2 \lambda} - {\cal F}_{2 \lambda_0}] \,
\kappa^2 \, P(x)/[x(1-x)] \quad , \eqno(4.6) $$

\noindent which is independent of the ultraviolet regularization.

$$ r_{\delta \mu}(x) = \left[ \tilde r_\delta(x) \right]^2 =
r^2_\delta(x) \, r^2_\delta(1-x) \quad . \eqno(4.7) $$

\noindent To find a more convenient notation for $\mu^2_\lambda$, the
arbitrary, possibly $\delta$-dependent number $\mu^2_{0\delta}$ can be
replaced by an expression that follows from a second order perturbative 
result for a single effective gluon mass correction obtained from
the eigenvalue equation for $H_\lambda(a_\lambda)$.  This expression
introduces a new parameter $\mu^2_\delta$ that would equal the second
order result for physical gluon mass squared, if such gluons existed.
With the form factor $f$ of Eq.  (3.15), one obtains

$$ \mu^2_\lambda = \mu^2_\delta + {g^2 \over (4\pi)^2 } \int_0^1 dx \,
r_{\delta \mu}(x) P(x) \int_0^\infty dz \, \exp{[-2z^2/\lambda^4]}
\quad , \eqno(4.8) $$

\noindent while the counterterm mass is

$$ \mu^2_\infty = \mu^2_\delta + {g^2 \over (4\pi)^2 } \int_0^1 dx \,
r_{\delta \mu}(x) P(x) \int_0^\infty dz \, \exp{[-4z x(1-x)/\Delta^2]}
\quad . \eqno(4.9) $$

\noindent Inclusion of $n_f$ flavors of quarks with bare masses $m_f$,
produces an additional term in the integrand in Eq.  (4.8), equal 
$\sum_{f=1}^{n_f} Q_f(x) \exp{[-2 z_f^2/\lambda^4] }$  with $Q_f(x) = 1-
2x(1-x) + 2m_f^2/z_f$ resembling the gluon splitting function into a
pair of massive quarks of flavor $f$, and $z_f = z + m_f^2/[x(1-x)]$.
Quarks would also add $\sum_{f=1}^{n_f} Q_f(x)$ to $P(x)$ in Eq.
(4.9).  Note that for convergence of the integral over $x$ in the
effective mass term (4.8), it is sufficient that the regulating
function behaves as $r_{\delta}(x) \sim x^\epsilon $ with $\epsilon >
0$.  In the counterterm (4.9), $r_{\delta}(x)$ has to vanish at small
$x$ at least as fast as $x^{1/2 + \epsilon}$.

Having the result for $\beta_{\lambda 11}$, one replaces the bare $a$
with effective $a_\lambda$.  The second order effective gluon mass
term is

$$ H_{(2)11} = \sum_{\sigma c} \int [k] {\mu_\lambda^2 \over k^+}
a^\dagger_{\lambda k\sigma c}a_{\lambda k\sigma c} \quad . \eqno(4.10)
$$

\noindent Other terms of order $g^2$ are derived following the same
path, but they do not require ultraviolet counterterms for
regularization adopted in Section 2.

Of main interest to this work is the term $\gamma_{\lambda 21}$ whose
calculation involves $\tau_1$, $\beta_{11}$, $\beta_{22}$ and
$\beta_{31}$.  Details of the calculation are described in Appendices
B and C. Eq.  (B1) for $\gamma_{\lambda 21}$ has the following
structure, cf.  Eq.  (3.11),

$$ \gamma_{\lambda 21} = \gamma_{\infty 21} + \int dx \,
d^2\kappa^\perp \, \left[ F_\lambda C + D \right] \, R_\Delta \quad
. \eqno(4.11)$$

\noindent $F_\lambda$ denotes form factors that multiply terms $C$,
which means that momentum integrals in terms $C$ cannot produce
dependence on the ultraviolet regularization factors $R_\Delta$ in the
limit $\Delta \rightarrow \infty$.  The first step in calculating
$\gamma_{\lambda 21}$ is to evaluate terms $\int D R_\Delta$ and find
their dependence on $R_\Delta$, to construct the counterterms
$\gamma_{\infty 21}$.  This step is carried out in Appendix C,
including $n_f$ flavors of quarks, with the result, see Eqs.  (B9) and
(C18), that

$$ \gamma_{\infty 21} = \sum_{123}\int[123] \not\!\delta(k_1 + k_2 -
k_3)\,\, { g^3 \over 16 \pi^3} \,\, \gamma_\infty \,\, a^\dagger_1
a^\dagger_2 a_3 \quad , \eqno(4.12) $$

\noindent where

$$ \gamma_\infty = Y_{123} { - \pi \over 3} \, \ln{ \Delta \over \mu}
\left\{ N_c\left[ 11 + h(x_1) \right] - 2 n_f \right\} +
\gamma_{finite} \quad , \eqno(4.13) $$

\noindent and

$$ h(x_1) = 6 \int_{x_1}^1 dx \,r_{\delta t}(x) \left[ {2 \over 1-x} +
{1 \over x-x_1} + {1 \over x} \right] - 9\, {\tilde r}_\delta(x_1)
\int_0^1 dx \,r_{\delta \mu}(x) \left[ {1\over x} + {1 \over 1-x}
\right] + (1 \leftrightarrow 2) \quad . \eqno(4.13a) $$

\noindent $\tilde r_\delta(x_1)$ is given by Eq.  (4.3), $r_{\delta
\mu}(x)$ by Eq.  (4.7), and $r_{\delta t}(x)$ by Eq.  (B4a).  The
counterterm part $\gamma_{finite}$ removes finite effects of
ultraviolet regularization.  It contains unknown functions of $x_1$
that multiply structures $Y_{12}$, $Y_{13}$, $Y_{23}$ in $Y_{123}$,
and the additional structure $Y_4$, see Eqs.  (C10) and (C15).
$\gamma_{finite}$ differs from canonical terms but it does not
influence the third order running of the Hamiltonian coupling constant.

Having found the counterterm, one obtains a finite answer for
$\gamma_{\lambda 21}$ in the limit $\Delta \rightarrow \infty$,

$$ \gamma_{\lambda 21} = \sum_{123}\int[123] \not\!\delta(p^\dagger -
p)\, \left[ W_{\lambda 12} Y_{12} + W_{\lambda 13} Y_{13} + W_{\lambda
23} Y_{23} + W_{\lambda 4} Y_4 \right] a^\dagger_1 a^\dagger_2 a_3
\quad, \eqno(4.14) $$

\noindent where the four coefficients $W_\lambda$ are functions of
$x_1$ and $\kappa^{\perp \,2}_{12}$.  The complete expression for
$\gamma_{\lambda 21}$ is given in Appendix B. The effective
three-gluon interaction term of order $g^3$ in $H_\lambda(a_\lambda)$
is $H_{(3)21} + H_{(3)12}$, where $H_{(3)12} = H_{(3)21}^\dagger$, and

$$ H_{(3)21} \,\, = \,\, f_\lambda \, {\cal U}_\lambda
\,\gamma_{\lambda 21} \, {\cal U}^\dagger_\lambda \quad . \eqno(4.15)
$$

\noindent The operator ${\cal U}_\lambda$ is unitary with accuracy to 
terms of order higher than 3rd and its action is equivalent here to 
changing $a$s to $a_\lambda$s as in lower order terms.

\vskip.3in {\bf 5. Three-gluon vertex}
\vskip.1in

This Section describes the running coupling constant $g_\lambda$ that
appears in Hamiltonian interaction vertices written in terms of gluons
of width $\lambda$.  The word ``running'' means changing with
$\lambda$.  The coupling constant is defined through the strength of
the three-gluon vertex for some value of $x_1=x_0$ when
$\kappa^\perp_{12} \rightarrow 0$, in direct analogy to the coupling
constant in $\phi^3$ theory in 6 dimensions~\cite{phi3} or electric
charge in the Thomson limit in QED.  The effective three-gluon vertex
is a sum of $H_{(1+3)21}$ and $H^\dagger_{(1+3)21}$, where $H_{(1+3)}
= H_{(1)} + H_{(3)}$ and

$$ H_{(1+3)21} = \sum_{123}\int[123] \not\!\delta(p^\dagger - p)\,
f_\lambda \left[ V_{\lambda 12} Y_{12} + V_{\lambda 13} Y_{13} +
V_{\lambda 23} Y_{23} + W_{\lambda 4} Y_4 \right] a^\dagger_{\lambda
1} a^\dagger_{\lambda 2} a_{\lambda 3} \quad, \eqno(5.1) $$

\noindent with $ f_\lambda = \exp{ - [\kappa^{\perp \, 2}_{12}/(x_1
x_2 \lambda^2)]^2}$.  The effective vertex contains creation and
annihilation operators for effective gluons of width $\lambda$,
instead of bare ones in Eqs.  (2.12b) and (4.14).  This change is
reflected by the presence of the form factor $f_\lambda$, which
strongly suppresses emission of effective gluons with transverse
momentum larger than $x_1 x_2 \lambda$.  The vertex functions
$V_{\lambda 12}(x_1, \kappa^{\perp \, 2}_{12})$, $V_{\lambda 13}(x_1,
\kappa^{\perp \, 2}_{12})$ and $V_{\lambda 23}(x_1, \kappa^{\perp \,
2}_{12})$ are used below to evaluate the running coupling constant.
The fourth function, $W_{\lambda 4}(x_1, \kappa^{\perp \, 2}_{12})$,
multiplies $Y_4$ which is distinct from canonical structures.
$W_{\lambda 4}$ is independent of $\lambda$ in the limit
$\kappa^\perp_{12} \rightarrow 0$ and does not contribute to the
running coupling constant.

When $\kappa^\perp_{12} \rightarrow 0$, all three vertex functions
$V_{\lambda ij}$ for $ij = 12, 13, 23 $, vary with $\lambda$ equally.
Using results from Appendix D without quarks, one can introduce a
single function,

$$ W_\lambda(x) \, = \, V_{\lambda ij}(x,0^\perp) -
V_{\lambda_0 ij}(x,0^\perp) $$
$$ = - \,{g^3 \over 48 \pi^2}\, N_c \, \left[11 + h(x) \right]
\,\ln{\lambda \over \lambda_0} \quad ,\eqno(5.2) $$

\noindent to describe behavior of all three functions 
$V_{\lambda ij}(x,0^\perp) = g \tilde r_\delta(x) + W_\lambda(x)$ 
for finite $x$ and $\lambda$ in the limit $\delta
\rightarrow 0$.

$W_\lambda(x)$ depends on $x$ through $h(x)$.  The case $h(x) = 0$
corresponds to the standard asymptotic freedom result~\cite{PGW} in
the following sense.  At certain value of $\lambda=\lambda_0$, the
three potentially different numbers $g_{\lambda_0 ij} \equiv
V_{\lambda_0 ij}(x_0,0^\perp)$ can be set equal to a common value $g_0 =
g_{\lambda_0}$ by adjusting finite parts of the counterterm (4.13), to
match the gauge symmetry result from the bare canonical Hamiltonian of
QCD.  For $h(x)=0$, the choice of $x_0$ does not matter as long as
$\delta$ is negligible in comparison to $x_0$ and $1-x_0$.  Then, for
$\lambda \ne \lambda_0$ the three coupling constants remain equal,
$g_{\lambda ij} = V_{\lambda ij}(x_0,0^\perp) = g_\lambda = g_0 +
W_\lambda(x_0) + o(g_0^5)$, with $g=g_0$ in $W_\lambda(x_0)$, or

$$ g_{\lambda } \,= \,g_0 \, -\, {g_0^3 \over 48 \pi^2}\, 11 N_c
\,\ln{\lambda \over \lambda_0}\quad . \eqno(5.3) $$

\noindent Expressing $g_0$ by $g_\lambda$ in the differential Eq.
(3.10), or in its integrated forms, (3.11) or (5.3), in the
perturbative expansion up to third order in $g_\lambda$, one has

$$ \lambda {d \over d\lambda} g_\lambda\, =\,\beta_0 \, g_\lambda^3
\quad , \eqno(5.4) $$

\noindent with

$$ \beta_0 \, = \, - \, { 11 N_c\over 48 \pi^2} , \eqno(5.5) $$

\noindent which is equal to the $\beta$-function coefficient in
Feynman calculus in QCD.  Thus, if one identifies the effective
Hamiltonian form factor width parameter $\lambda$ with the
running momentum scale in Feynman diagrams, the standard result from
off-shell $S$-matrix calculus would be recovered in effective
Hamiltonians in the case $h(x) = 0$.

The function $h(x)$ is determined by the initial Hamiltonian and
small-$x$ regularization factor $r_\delta$ in it through Eq.  (4.13a).
In the limit $\delta \rightarrow 0$, for $r_\delta$ of Eq.  (2.8),

$$ h(x) \, = \, 12\left[ 3 + {1-x - x^2 \over (1-x)(1-2x)} \ln{x} +
{(1-x)^2 -x \over x(1-2x) } \ln{(1-x)} \right], \eqno(5.6) $$

\noindent and in the case of $r_\delta$ from Eq.  (2.9),

$$ h(x) \, = \, 12 \ln{ min(x,1-x)} . \eqno(5.7) $$

\noindent These two cases are shown in Fig. 1 along with the case c)
that corresponds to $h(x) = 0$. The vertex function $V(x)$ in Fig. 1,
is defined by

$$ g_0 V(x) = g_0 + W_\lambda(x) \quad , \eqno(5.8)$$

\noindent and plotted using $\alpha_0 = g_0^2/(4\pi) = 0.1$ for
$\lambda_0=100$ GeV, and $N_c=3$.
In part d) of Fig. 1, the value of $V(1/2)$ is plotted against
$\lambda$.   The sharp cutoff case b) of $\theta(x-\delta)$ is visibly
different from c)  that corresponds to $h(x_1)=0$, but the continuous
case a) of $x/(x+\delta)$  differs from c) only by 8\%.  By
extrapolation, the results a) and b) suggest that there may exist a
regularization factor $r_\delta(x)$ that matches the case c).  The
mathematical structure of Eq.  (4.13a) and Euclidean integrals in
Feynman diagrams with dimensional regularization~\cite{Peskin} both
hint at the power-law functions $r_\delta(x) = x^\delta$.
Inspection shows that in the limit $\delta \rightarrow 0$,

\vskip.1in \centerline{\psfig{figure=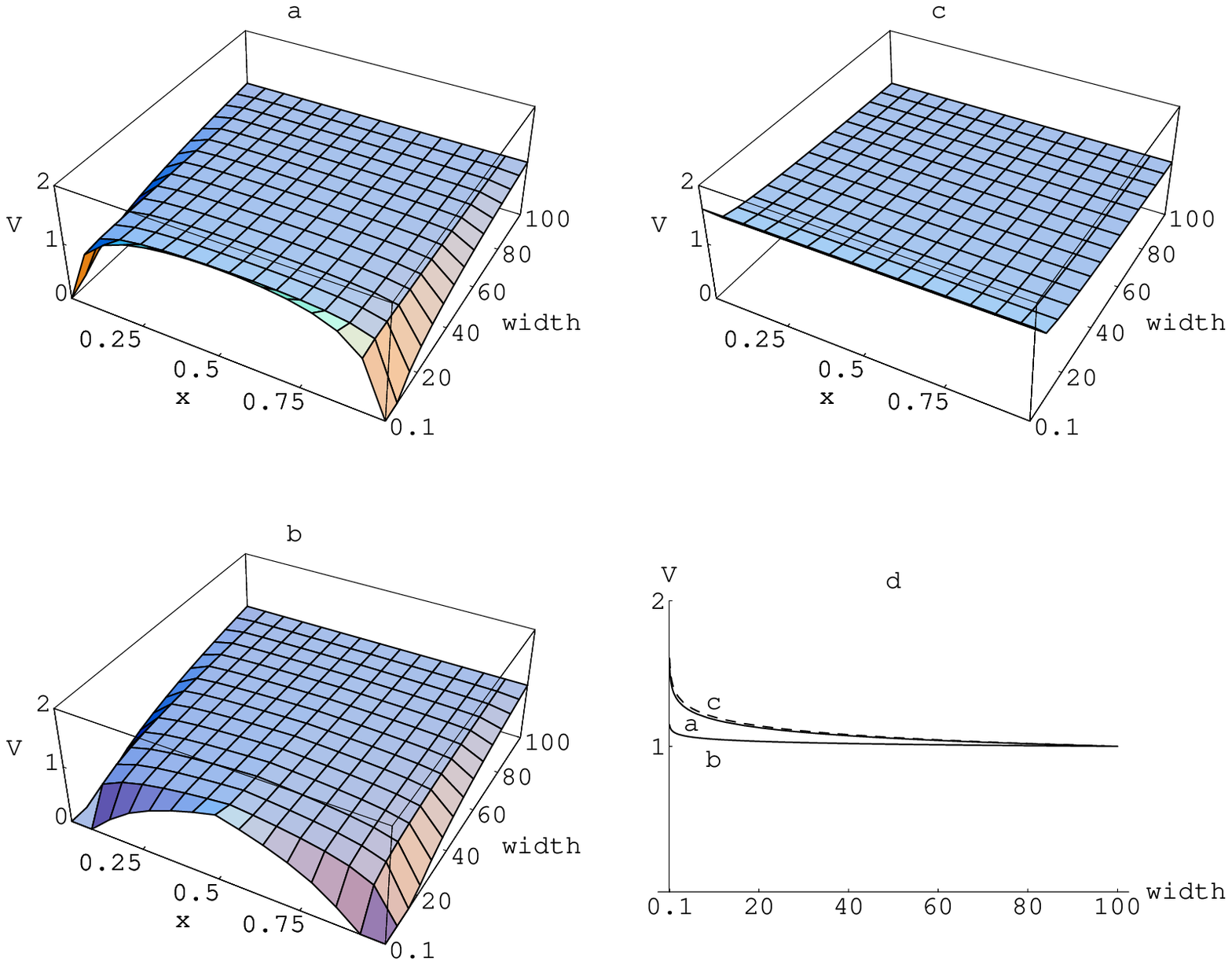}}
\begin{center}
\parbox{15cm}{\small {\bf Fig. 1} Change of the effective gluon vertex
function $V(x)$ from Eq.  (5.8) with the width $\lambda$ varying from
$\lambda_0 = 100$ GeV down to 100 MeV, for three different small-$x$
regularization functions:  a) $r_\delta = x/(x+\delta)$, b) $r_\delta
= \theta(x-\delta)$ and c) $r_\delta = x^\delta \,
\theta(x-\epsilon)$.  Part d) shows dependence of $V(1/2)$ on
$\lambda$ for the three cases, correspondingly.  The case c, dashed
line, matches the QCD running coupling constant result obtained from
Feynman diagrams.  Note the dynamical suppression of the effective
gluon coupling for extreme values of $x$ in cases a) and b).  See the
text for details.}
\end{center}

$$ r_\delta(x) \, = \, x^\delta \, \theta(x-\epsilon) \eqno(5.9) $$

\noindent leads to $ h(x_1) = 0$ for all fixed values of $x_1$ between
0 and 1, as long as $\epsilon/\delta \rightarrow 0$.  This result sets
a path for developing a connection between the Hamiltonian dynamics of
effective gluons in the light-front Fock space, and Feynman diagrams
for Green's functions in Lagrangian calculus~\cite{PGW}.

The second scale $\epsilon$ in Eq.  (5.9) is required to regulate
linear small-$x$ divergences such as in the mass counterterms in Eq.
(4.9).  Details of $r_\delta(x)$ at $x \sim \epsilon$ are not
important for point-wise convergence of $h(x_1)$ to 0, and a whole
class of regularizations with a second scale $\epsilon \ll \delta$
give the same result.  Ultraviolet regularizations using invariant
mass, soften small-$x$ divergences in mass counterterms so that the
second scale $\epsilon$ appears unnecessary, but such regularizations
mix small-$x$ effects with large $\kappa^\perp$ divergences, see
Appendix E. In evaluation of $h(x_1)$, factorization of the
ultraviolet renormalization group flow in $\lambda$ from small-$x$
effects requires $\delta \rightarrow 0$.  The small-$x$ behavior of
other terms in this limit has to be compared with the size of $g$ to
discuss validity of the perturbative analysis.  The warning is
warranted by the fact that in Fig. 1, where the value of $\alpha \sim
0.1$ for $\lambda$ on the order of 100 GeV is taken from phenomenology
based on Feynman diagrams, $g_0 \sim 1.1 > 1$.

Fig. 1 shows that small-$x$ regularizations of Eq.  (2.8) and (2.9)
lead to suppression of effective gluon interactions when $x_1$ moves
away from 1/2 for $\kappa^\perp_{12} \rightarrow 0$, where $f_\lambda
\rightarrow 1$.  They also slow down the rate of growth of the
coupling constant with lowering $\lambda$.  Similar results follow for
$f_\lambda$ depending on $p^-$ instead of ${\cal M}^2$.  $\lambda^2$
is replaced by $k_3^+ \lambda$ and one looses boost invariance, but in
the ratio $(\lambda_1 k_3^+)/(\lambda_2 k_3^+)$ the dependence on
$k_3^+$ cancels out.

Thorn~\cite{Thorn} calculated the diverging part of a 4-point gluon
Green's function using $A^+=0$ gauge and integrating Feynman diagrams
first over $k^-$.  Using sharp cutoffs $k^+ > \epsilon^+$ and
$|k^\perp| \le \Lambda$ in the remaining integrals, Thorn found that
to understand asymptotic freedom one has to include direct and crossed
box diagrams that cancel skew $p^+$-dependent terms from the
three-point function whose own cutoff dependence had opposite sign to
asymptotic freedom.  Lepage and Brodsky~\cite{LB} developed a whole
formalism for hard exclusive processes involving hadrons.
Perry~\cite{Perryaf} used their rules in old fashioned perturbation
theory to obtain the ultraviolet diverging part of o set of third
order terms in the quark-gluon-quark off-energy-shell vertex function,
with similar cutoffs to Thorn's.  In the limit of ultraviolet cutoffs
being sent to infinity, Perry found asymptotic freedom in the cutoff
dependence of the quark-gluon vertex function.  Irrelevant parts
diverged as functions of the small-$p^+$ cutoff and depended on the
off-shell energy parameter, but Perry reported that they could be made
small by using an invariant mass cutoff.  In calculations of the
Green's functions, small energy denominators require an extra step of
setting a lower bound on transverse momenta and factorization of the
contributions from below.  Such contributions complicate issues that
arise in the introduction of the renormalization scale at small scales, 
and gluon coupling must be brought under control~\cite{BLM}.  The 
dependence on the energy-shell parameter involves Hamiltonian eigenvalues, 
which depend on solution of the hadronic binding problem.

Effective Hamiltonians are calculated differently from Green's
functions.  Asymptotic freedom appears in dependence of the
Hamiltonian vertices on width $\lambda$.  The width may be finite and
adjusted to the scale of processes one is interested in, independently
of the diverging ultraviolet cutoffs. ${\cal U}_\lambda$ is unitary
and  the procedure involves no wave function renormalization, in
distinction  from the standard renormalization group concept, and no
Ward identity is  used.  $\lambda$ effectively limits energy
denominators from below, so  that issues of binding are clearly
separated from logarithmic evolution  of effective Hamiltonians.  No
off-shell energy parameter is introduced.   In summary, no need arises
to rely on the arbitrary ultraviolet  regularization, no extra lower
bound on momenta is needed, and no off-shell energy parameter is
introduced.  The notion of effective constituents is then a natural
candidate for the phenomenology of hadronic wave functions~\cite{BH}
to be put on the firm ground of Hamiltonian quantum mechanics, with an
open path to make connection with diagrammatic
techniques~\cite{Bassetto} for scattering amplitudes.

Although the renormalized Hamiltonian vertices display a tendency to
decouple dynamics of effective gluons from small-$x$ region, many $x$
dependent terms completely drop out from the third order running of
the coupling constant. These terms require careful investigation since
vacuum effects may enter through the small-$x$ region~\cite{Susskind}.
Once these terms are calculated in detail using the perturbative procedure, 
they can be subsequently studied in the Schr\"odinger equations with effective
Hamiltonians beyond perturbation theory.  In particular, variational
studies could aim at understanding gluon condensation and spontaneous
chiral symmetry breaking, including the cases where the number of
flavors  approaches the critical value at which asymptotic freedom
goes  away~\cite{Appelquist}.  The small-$x$ regularization clearly
interferes in this transition, see Eq.  (4.13).

\vskip.3in {\bf 6. Conclusion}
\vskip.1in

Coefficients of products of creation and annihilation operators for
effective gluons of width $\lambda$ in renormalized light-front QCD
Hamiltonian, contain vertex form factors $f_\lambda$ and vary with
$\lambda$ in a perturbatively calculable way. Obtained in third order
perturbation theory, the effective three-gluon  vertex with vanishing
transverse momentum, is a linear function of $\ln \lambda$. The
coefficient of $\ln \lambda$ matches the coefficient of $\ln Q$  that
appears in the running coupling constant dependence on the running
scale $Q$ in Lagrangian calculus for Green's functions. This happens
for small-$x$ regularizations of the type $r_\delta(x) = x^\delta
\theta(x-\epsilon)$ in the limit $\delta \rightarrow 0$ for
$\epsilon/\delta \rightarrow 0$.  For other regularizations, such as
$r_\delta(x) = x/(x+\delta)$ or $r_\delta(x) = \theta(x-\delta)$, the
coefficient of $\ln \lambda$ contains an additive term $h(x)$, defined
in Eq.  (4.13a) and described in Section 5. $h(x)$ suppresses
interactions  at small $x$ in addition to the suppression implied by
the vertex form  factors $f_\lambda$.

In the effective particle light-front Fock space basis in gauge
$A^+=0$, asymptotic freedom is a consequence of that a single
effective gluon contains a pair of bare gluons.  This component
amplifies the strength with which effective gluons can split into
effective gluons  when $\lambda$ gets smaller. The mechanism is  the
same as in the case of scalar particles in 6 dimensions~\cite{phi3},
although the source of $\ln \lambda$ is different.  Namely, in  scalar
theory it is the integration over additional transverse  momentum
dimensions, while in QCD it is the transverse momentum  factors in
gluon polarization vectors.  In perturbative description of processes
characterized by a physical momentum scale $Q$, using
$H_\lambda(a_\lambda)$, there will appear powers of $g_\lambda$ and
$\ln{Q/\lambda}$.  For $Q/\lambda=1$, $\ln{Q/\lambda}=0$, so that the
theoretical predictions will have the form of a series in powers of
the asymptotically free running coupling constant $g_Q$.

The dynamics of effective particles includes binding, described
through Schr\"odinger equation with $H_\lambda(a_\lambda)$.  Initial
studies of some simplified bound state eigenvalue problems for
light-front QCD Hamiltonian matrices, have recently been carried out 
for heavy quarkonia~\cite{BPW} and gluonium 
states~\cite{AllenPerrygluonium}, reporting reasonable results.
Although the effective particle approach described here is different,
the bound state equations resulting from the second order perturbation
theory may be similar.  So, the initial studies suggest that the
effective particle approach should be tested in application to
bound states of constituent quarks and gluons.

Besides the bound state dynamics of low energy constituents in the
hadronic rest frame, the boost-invariant effective particle approach
provides a theoretical tool for studies of layers of hadronic
structure in other frames of reference, including the infinite
momentum  frame.  The theory is simplest, and in lowest order of the
same type as  for quarks and gluons, in the case of electron-hadron
interaction through  one-photon exchange. In the case of massive
vector bosons, there exist  three instead of only two polarization
states and the choice of gauge  $A^+=0$ is not directly available. A
scheme to attempt in the more  complicated theory was formulated by
Soper~\cite{Soper} in massive QED,  but the required renormalization
procedure for Hamiltonians  with massive gauge bosons is completely
undeveloped in comparison to the  current status of renormalization in
Lagrangian calculus~\cite{tHW}.

Physical processes that involve one photon exchange have amplitudes
proportional to $e^2$.  To obtain an amplitude for a simplest
scattering event with large momentum transfer that involves strong
interactions, it is sufficient to calculate the strong subprocess
only to order $g^2$ and study cross-sections to order $e^4 g^2$.   The
complete initial Hamiltonian contains four terms,

$$ H = H_{QED} + H_{e q \gamma} + H_{QCD} + X \quad , \eqno(6.1) $$

\noindent where the second term includes couplings of quarks to
photons and instantaneous electromagnetic coupling of quarks and
electrons.  The transformation ${\cal U}_\lambda$ of Eq.  (3.1) can be
calculated from the trajectory of $H_{\lambda \, QCD}$ alone, together
with the required counterterms in $X$.  Then, the Hamiltonians $H_{e q
\gamma}$ can be re-written in terms of quarks and gluons of width
$\lambda$.  This step produces dependence on regularization, which
should be removed by additional counterterms in $X$.  QED degrees of
freedom are not changed.  The resulting effective Hamiltonian dynamics
can be then tested for covariance using perturbative expansion for the
amplitude $e^+e^- \rightarrow hadrons$ including terms order $e^2$,
$e^2 g$ and $e^2 g^2$, that contribute to the cross section through
orders $e^4$ and $e^4 g^2$.

Such tests are of interest since the renormalization group for
particles produces light-front Hamiltonians without reference to the
vacuum structure and scattering amplitudes, and wave function
renormalization and Ward identities do not enter the Hamiltonian. In a
much simplified  model without gauge symmetry, adjustment of finite
parts of counterterms led to covariant scattering with proper
threshold behavior~\cite{MW}. It should be verified in the Hamiltonian
approach  based on Eq. (6.1) if QCD binding effects known in order
$g^2$ interfere  with obtaining covariant results. Genuinely
perturbative fourth-order  test calculations in QED should verify if
in the limit $\delta
\rightarrow 0$ the perturbative Hamiltonian approach is able to produce 
fully covariant results for observables through adjustment of finite
parts of the ultraviolet counterterms that contain otherwise
undetermined  functions of $x$.

\vskip.3in {\bf Appendix A: Details of the initial Hamiltonian}
\vskip.1in

The initial QCD Hamiltonian for gluons in Eq.  (2.11) contains the
following terms.

$$ H_{A^2} = \sum_{\sigma c} \int [k] {k^{\perp \, 2} \over k^+}
a^\dagger_{k\sigma c}a_{k\sigma c} \quad. \eqno(A1) $$

$$ H_{A^3} = \sum_{123}\int[123] \not\!\delta(p^\dagger - p)\,\tilde
r_{\Delta\delta}(3,1) \left[g\,Y_{123}\, a^\dagger_1 a^\dagger_2 a_3 +
g\,Y_{123}^*\, a^\dagger_3 a_2 a_1 \right] , \eqno(A2) $$
\noindent where

$$ Y_{123} = i f^{c_1 c_2 c_3} \left[ \varepsilon_1^*\varepsilon_2^*
\cdot \varepsilon_3\kappa - \varepsilon_1^*\varepsilon_3 \cdot
\varepsilon_2^*\kappa {1\over x_{2/3}} - \varepsilon_2^*\varepsilon_3
\cdot \varepsilon_1^*\kappa {1\over x_{1/3}} \right], \eqno(A2a) $$

\noindent with $\varepsilon \equiv \varepsilon^\perp$ and $ \kappa
\equiv \kappa^\perp_{1/3}$.

$$ H_{A^4} = \sum_{1234}\int[1234] \not\!\delta(p^\dagger - p)\,{g^2
\over 4}\, \left[ \Xi_{A^4 \, 1234} a^\dagger_1 a^\dagger_2
a^\dagger_3 a_4 + X_{A^4\,1234} a^\dagger_1 a^\dagger_2 a_3 a_4 +
\Xi^*_{A^4\,1234} a^\dagger_4 a_3 a_2 a_1 \right] \, . \eqno(A3) $$

$ \Xi_{A^4 \, 1234} = {2\over 3}[$
\parbox[t]{9.9cm}{ $ \tilde r_{1+2,1} \tilde r_{4,3}\,
(\varepsilon_1^*\varepsilon_3^* \cdot \varepsilon_2^*\varepsilon_4 -
\varepsilon_1^*\varepsilon_4 \cdot \varepsilon_2^*\varepsilon_3^*)\,
f^{a c_1 c_2}f^{a c_3 c_4} + $
\vskip.1in $ \tilde r_{1+3,1} \tilde r_{4,2}\,
(\varepsilon_1^*\varepsilon_2^* \cdot \varepsilon_3^*\varepsilon_4 -
\varepsilon_1^*\varepsilon_4 \cdot \varepsilon_2^*\varepsilon_3^*)\,
f^{a c_1 c_3}f^{a c_2 c_4} + $
\vskip.1in $ \tilde r_{3+2,3} \tilde r_{4,1}\,
(\varepsilon_1^*\varepsilon_3^* \cdot \varepsilon_2^*\varepsilon_4 -
\varepsilon_3^*\varepsilon_4 \cdot \varepsilon_2^*\varepsilon_1^*)\,
f^{a c_3 c_2}f^{a c_1 c_4}] \, . $ }
\parbox[t]{4cm}{\vskip1pt
\begin{picture}(80,50)(0,0)
\multiput(28,37)(-.3,.3){65}{\rule{.3pt}{.3pt}}
\multiput(18,47)(-.3,-.3){32}{\rule{.3pt}{.3pt}}
\multiput(13,52)(-.3,-.3){16}{\rule{.3pt}{.3pt}} \put(0,55){\tiny 1}
\put(0,45){\tiny 2} \put(0,35){\tiny 3} \put(30,35){\tiny 4}
\multiput(78,22)(-.3,.3){65}{\rule{.3pt}{.3pt}}
\multiput(70,30)(-.3,-.0){36}{\rule{.3pt}{.3pt}}
\multiput(65,35)(-.1,-.3){13}{\rule{.3pt}{.3pt}}
\multiput(60,20)(.1,.3){28}{\rule{.3pt}{.3pt}} \put(50,40){\tiny 1}
\put(50,30){\tiny 2} \put(50,20){\tiny 3} \put(80,20){\tiny 4}
\multiput(28,7)(-.3,.3){63}{\rule{.3pt}{.3pt}}
\multiput(18,17)(-.3,-.3){32}{\rule{.3pt}{.3pt}}
\multiput(13,12)(-.3,.3){16}{\rule{.3pt}{.3pt}} \put(0,25){\tiny 1}
\put(0,15){\tiny 2} \put(0,5){\tiny 3} \put(30,5){\tiny 4}
\end{picture}}
\vskip.1in $ X_{A^4 \, 1234} =~~~$
\parbox[t]{10cm}{ $ \tilde r_{1+2,1} \tilde r_{3+4,3}\,
(\varepsilon_1^*\varepsilon_3 \cdot \varepsilon_2^*\varepsilon_4 -
\varepsilon_1^*\varepsilon_4 \cdot \varepsilon_2^*\varepsilon_3)\,
f^{a c_1 c_2}f^{a c_3 c_4} + $
\vskip.1in $ [\tilde r_{3,1} \tilde r_{2,4}+\tilde r_{1,3} \tilde
r_{4,2}]\, (\varepsilon_1^*\varepsilon_2^* \cdot \varepsilon_3
\varepsilon_4 - \varepsilon_1^*\varepsilon_4 \cdot
\varepsilon_2^*\varepsilon_3)\, f^{a c_1 c_3}f^{a c_2 c_4} + $
\vskip.1in $ [\tilde r_{3,2} \tilde r_{1,4} + \tilde r_{2,3} \tilde
r_{4,1}] \, (\varepsilon_1^*\varepsilon_2^* \cdot
\varepsilon_3^*\varepsilon_4 - \varepsilon_1^*\varepsilon_3 \cdot
\varepsilon_2^*\varepsilon_4)\, f^{a c_1 c_4}f^{a c_2 c_3} \, . $ }
\parbox[t]{3.5cm}{\vskip1pt
\begin{picture}(70,55)(0,0)
\put(25,60){\tiny 1} \put(25,45){\tiny 2} \put(55,60){\tiny 3}
\put(55,45){\tiny 4} \multiput(38,54)(.3,0){27}{\rule{.3pt}{.3pt}}
\multiput(38,54)(-.3,.3){27}{\rule{.3pt}{.3pt}}
\multiput(38,54)(-.3,-.3){27}{\rule{.3pt}{.3pt}}
\multiput(46,54)(.3,.3){27}{\rule{.3pt}{.3pt}}
\multiput(46,54)(.3,-.3){27}{\rule{.3pt}{.3pt}}
\put(5,40){\tiny 1} \put(5,25){\tiny 2} \put(35,40){\tiny 3}
\put(35,25){\tiny 4} \multiput(10,42)(.3,0){80}{\rule{.3pt}{.3pt}}
\multiput(10,27)(.3,0){80}{\rule{.3pt}{.3pt}}
\multiput(18,27)(.2,.3){50}{\rule{.3pt}{.3pt}}
\put(5,20){\tiny 1} \put(5,5){\tiny 2} \put(35,20){\tiny 3}
\put(35,5){\tiny 4} \multiput(28,22)(.3,0){24}{\rule{.3pt}{.3pt}}
\multiput(28,22)(-.33,-.3){50}{\rule{.3pt}{.3pt}}
\multiput(18,7)(.3,0){50}{\rule{.3pt}{.3pt}}
\multiput(18,7)(.2,.3){50}{\rule{.3pt}{.3pt}}
\multiput(18,7)(-.15,.3){50}{\rule{.3pt}{.3pt}}
\put(60,40){\tiny 1} \put(60,25){\tiny 2} \put(90,40){\tiny 3}
\put(90,25){\tiny 4} \multiput(65,42)(.3,0){80}{\rule{.3pt}{.3pt}}
\multiput(65,27)(.3,0){80}{\rule{.3pt}{.3pt}}
\multiput(73,42)(.2,-.3){50}{\rule{.3pt}{.3pt}}
\put(60,20){\tiny 1} \put(60,5){\tiny 2} \put(90,20){\tiny 3}
\put(90,5){\tiny 4} \multiput(73,22)(.3,0){48}{\rule{.3pt}{.3pt}}
\multiput(83,7)(.3,0){24}{\rule{.3pt}{.3pt}}
\multiput(73,22)(.2,-.3){50}{\rule{.3pt}{.3pt}}
\multiput(83,7)(-.33,.3){50}{\rule{.3pt}{.3pt}}
\multiput(73,22)(-.15,-.3){50}{\rule{.3pt}{.3pt}}
\end{picture}}

\vskip.1in

$$ H_{[\partial A A]^2} = \sum_{1234}\int[1234] \not\!\delta(p^\dagger
- p)\,g^2\, \left[ \left( \Xi_{[\partial A A]^2 \, 1234} a^\dagger_1
a^\dagger_2 a^\dagger_3 a_4 + h.c. \right) + X_{[\partial A
A]^2\,1234} a^\dagger_1 a^\dagger_2 a_3 a_4 \right] \,.  \eqno(A4) $$
\vskip.1in $ \Xi_{[\partial A A]^2\, 1234} = - {1\over 6}[$
\parbox[t]{9cm}{ $ \tilde r_{1+2,1} \tilde r_{4,3}\,
\varepsilon_1^*\varepsilon_2^* \cdot \varepsilon_3^*\varepsilon_4\,
{(x_1 - x_2)(x_3+x_4)\over (x_1 + x_2)^2}\, f^{a c_1 c_2}f^{a c_3 c_4}
+ $
\vskip.05in $ \tilde r_{1+3,1} \tilde r_{4,2}\,
\varepsilon_1^*\varepsilon_3^* \cdot \varepsilon_2^*\varepsilon_4\,
{(x_1 - x_3)(x_2+x_4)\over (x_1 + x_3)^2}\, f^{a c_1 c_3}f^{a c_2 c_4}
+ $
\vskip.05in $ \tilde r_{3+2,3} \tilde r_{4,1}\,
\varepsilon_3^*\varepsilon_2^* \cdot \varepsilon_1^*\varepsilon_4\,
{(x_3 - x_2)(x_1+x_4)\over (x_3 + x_2)^2}\, f^{a c_3 c_2}f^{a c_1
c_4}] \, . $ }
\parbox[t]{4cm}{\vskip1pt
\begin{picture}(80,50)(0,0)
\multiput(28,37)(-.3,.3){65}{\rule{.3pt}{.3pt}}
\multiput(18,47)(-.3,-.3){32}{\rule{.3pt}{.3pt}}
\multiput(13,52)(-.3,-.3){16}{\rule{.3pt}{.3pt}} \put(0,55){\tiny 1}
\put(0,45){\tiny 2} \put(0,35){\tiny 3} \put(30,35){\tiny 4}
\multiput(78,22)(-.3,.3){65}{\rule{.3pt}{.3pt}}
\multiput(70,30)(-.3,-.0){36}{\rule{.3pt}{.3pt}}
\multiput(65,35)(-.1,-.3){13}{\rule{.3pt}{.3pt}}
\multiput(60,20)(.1,.3){28}{\rule{.3pt}{.3pt}} \put(50,40){\tiny 1}
\put(50,30){\tiny 2} \put(50,20){\tiny 3} \put(80,20){\tiny 4}
\multiput(28,7)(-.3,.3){63}{\rule{.3pt}{.3pt}}
\multiput(18,17)(-.3,-.3){32}{\rule{.3pt}{.3pt}}
\multiput(13,12)(-.3,.3){16}{\rule{.3pt}{.3pt}} \put(0,25){\tiny 1}
\put(0,15){\tiny 2} \put(0,5){\tiny 3} \put(30,5){\tiny 4}
\end{picture}}
\vskip.1in $ X_{[\partial A A]^2\, 1234}={1\over 4}[$
\parbox[t]{10.2cm}{ $ \tilde r_{1+2,1} \tilde r_{3+4,3}\,
\varepsilon_1^*\varepsilon_2^* \cdot \varepsilon_3 \varepsilon_4\,
{(x_1 - x_2)(x_3 - x_4)\over (x_1 + x_2)^2}\, f^{a c_1 c_2}f^{a c_3
c_4} - $
\vskip.05in $ [\tilde r_{3,1} \tilde r_{2,4}+\tilde r_{1,3} \tilde
r_{4,2}]\, \varepsilon_1^*\varepsilon_3 \cdot \varepsilon_2^*
\varepsilon_4\, {(x_1 + x_3)(x_2 + x_4)\over (x_2 - x_4)^2}\, f^{a c_1
c_3}f^{a c_2 c_4} - $
\vskip.05in $ [\tilde r_{3,2} \tilde r_{1,4} + \tilde r_{2,3} \tilde
r_{4,1}] \, \varepsilon_1^*\varepsilon_4 \cdot
\varepsilon_2^*\varepsilon_4\, {(x_2 + x_3)(x_1+x_4)\over (x_1 -
x_4)^2}\, f^{a c_1 c_4}f^{a c_2 c_3}] \, . $ }
\parbox[t]{3cm}{\vskip1pt
\begin{picture}(70,55)(0,0)
\put(25,60){\tiny 1} \put(25,45){\tiny 2} \put(55,60){\tiny 3}
\put(55,45){\tiny 4} \multiput(38,54)(.3,0){27}{\rule{.3pt}{.3pt}}
\multiput(38,54)(-.3,.3){27}{\rule{.3pt}{.3pt}}
\multiput(38,54)(-.3,-.3){27}{\rule{.3pt}{.3pt}}
\multiput(46,54)(.3,.3){27}{\rule{.3pt}{.3pt}}
\multiput(46,54)(.3,-.3){27}{\rule{.3pt}{.3pt}}
\put(5,40){\tiny 1} \put(5,25){\tiny 2} \put(35,40){\tiny 3}
\put(35,25){\tiny 4} \multiput(10,42)(.3,0){80}{\rule{.3pt}{.3pt}}
\multiput(10,27)(.3,0){80}{\rule{.3pt}{.3pt}}
\multiput(18,27)(.2,.3){50}{\rule{.3pt}{.3pt}}
\put(5,20){\tiny 1} \put(5,5){\tiny 2} \put(35,20){\tiny 3}
\put(35,5){\tiny 4} \multiput(28,22)(.3,0){24}{\rule{.3pt}{.3pt}}
\multiput(28,22)(-.33,-.3){50}{\rule{.3pt}{.3pt}}
\multiput(18,7)(.3,0){50}{\rule{.3pt}{.3pt}}
\multiput(18,7)(.2,.3){50}{\rule{.3pt}{.3pt}}
\multiput(18,7)(-.15,.3){50}{\rule{.3pt}{.3pt}}
\put(50,40){\tiny 1} \put(50,25){\tiny 2} \put(80,40){\tiny 3}
\put(80,25){\tiny 4} \multiput(55,42)(.3,0){80}{\rule{.3pt}{.3pt}}
\multiput(55,27)(.3,0){80}{\rule{.3pt}{.3pt}}
\multiput(63,42)(.2,-.3){50}{\rule{.3pt}{.3pt}}
\put(50,20){\tiny 1} \put(50,5){\tiny 2} \put(80,20){\tiny 3}
\put(80,5){\tiny 4} \multiput(63,22)(.3,0){48}{\rule{.3pt}{.3pt}}
\multiput(73,7)(.3,0){24}{\rule{.3pt}{.3pt}}
\multiput(63,22)(.2,-.3){50}{\rule{.3pt}{.3pt}}
\multiput(73,7)(-.33,.3){50}{\rule{.3pt}{.3pt}}
\multiput(63,22)(-.15,-.3){50}{\rule{.3pt}{.3pt}}
\end{picture}}

\vskip.1in

\noindent $p^\dagger$ and $p$ are the total momenta of created and
annihilated particles, respectively. For the vertex with parent
momentum $p$ and daughter momenta $d$ and $p-d$, the regulating
function in Eq.  (A2) is $  \tilde r_{\Delta\delta}(p,d) =
r_{\Delta\delta}(p,d) \,r_{\Delta\delta}(p,p-d)$, where $
r_{\Delta\delta}(p,d) = r_\Delta(\kappa^{\perp \, 2}_{d/p})
\,r_\delta(x_{d/p})\,\theta(x_{d/p}) \quad$, with $ x_{d/p} = d^+/p^+
\equiv x_d/x_p $ and $\kappa^\perp_{d/p} = d^\perp - x_{d/p} p^\perp
$.  Eqs.  (2.6) to (2.9) complete the definitions of $\tilde
r_{\Delta\delta}(p,d)$. Also, $ \tilde r_{p,d} \equiv \tilde
r_{\Delta\delta}(p,d)$.  The momentum integration measures contain the
same factors as in Eq.  (2.4) for each indicated particle.

\vskip.3in
{\bf Appendix B: Expression for $\gamma_{\lambda 21}$ }
\vskip.1in

Eq. (3.19) gives
\begin{eqnarray*}
 \gamma_{\lambda 21} &  = & {\cal F}_{3\lambda} \, 8 \left[
\alpha_{12} \alpha_{21} \alpha_{21}\right]_{21} + {\cal F}_{2\lambda}
\, 2 \left[ \beta_{\infty 22} \alpha_{21}\right]_{21} + {\cal
F}_{2\lambda} \, 2 \left[ \alpha_{12} \beta_{\infty 31} \right]_{21}
\\ & & \\ & + & \, {\cal F}_{3\lambda} \, 4 \left[ [\alpha_{12}
\alpha_{21}]_{11} \alpha_{21}\right]_{21} + {\cal F}_{2\lambda} \,
\left[ \alpha_{12} \beta_{\infty 31} \right]_{21}
+ {\cal F}_{2\lambda} \, 2 \left[ \beta_{\infty 11}
\alpha_{21}\right]_{21} \\ & & \\ & + &\, {\cal F}_{3\lambda} \, 2
\left[ \alpha_{21} [\alpha_{12} \alpha_{21}]_{11}\right]_{21} + {\cal
F}_{2\lambda} \, {1\over 2} \left[ \beta_{\infty 22} \alpha_{21}
\right]_{21} + {\cal F}_{2\lambda} \,
\left[ \alpha_{21} \beta_{\infty 11} \right]_{21}\\
& & \\ & + & \, \gamma_{\infty 21} \quad .
\end{eqnarray*}
$$ \eqno(B1) $$

\noindent The terms are grouped and ordered in one to one correspondence 
to Fig. 2, so that using labels from Fig.  2, Eq.  (B1) reads:

$$ \gamma_{21} = {\cal F}_{3} 8 { (a)} + {\cal F}_{2} 2 { (b)} + {\cal
F}_{2} 2 { (c)} + {\cal F}_{3} 4 { (d)} + {\cal F}_{2} { (e)} + {\cal
F}_{2} 2 { (f)} + {\cal F}_{3} 2 { (g)} + {\cal F}_{2} {1\over 2} {
(h)} + {\cal F}_{2} { (i)} + { (j)} \, . \eqno(B2) $$

\begin{center}
\parbox[t]{9cm}{
\begin{picture}(200,230)(0,70)

\setlength{\unitlength}{.5mm}
\thicklines

\multiput(20,190)(-.3,.15){40}{\rule{.8pt}{.8pt}}
\multiput(20,190)(-.3,-.6){40}{\rule{.8pt}{.8pt}}
\multiput(20,190)(.3,-.6){40}{\rule{.8pt}{.8pt}}
\multiput(20,175)(.3,0){25}{\rule{.8pt}{.8pt}}
\multiput(20,175)(-.3,0){25}{\rule{.8pt}{.8pt}}
\put(0,195){\small 1}
\put(10,180){\small 6}
\put(27,180){\small 7}
\put(18,169){\small 8}
\put(0,165){\small 2}
\put(35,165){\small 3}
\put(15,155){\bf $(a)$}

\multiput(100,190)(-.3,.15){40}{\rule{.8pt}{.8pt}}
\multiput(100,190)(-.3,-.6){40}{\rule{.8pt}{.8pt}}
\multiput(99.3,189)(-.3,-.6){23}{\rule{.9pt}{.9pt}}
\multiput(100,189)(-.3,-.6){23}{\rule{.9pt}{.9pt}}
\multiput(100,190)(.3,-.6){40}{\rule{.8pt}{.8pt}}
\multiput(100,175)(.3,0){25}{\rule{.8pt}{.8pt}}
\multiput(100,175)(-.3,0){25}{\rule{.8pt}{.8pt}}
\put(80,195){\small $x_1$}
\put(80,182){\small $x$-$x_1$}
\put(107,182){\small $x$}
\put(95,169){\small $1$-$x$}
\put(80,165){\small $x_2$}
\put(115,165){\small $x_3$}
\multiput(94,183)(.6,-.3){8}{\rule{.9pt}{.9pt}}
\multiput(94,183.6)(.6,-.3){8}{\rule{.9pt}{.9pt}}
\put(95,155){\bf $(b)$}

\multiput(180,190)(-.3,.15){40}{\rule{.8pt}{.8pt}}
\multiput(180,190)(-.3,-.6){40}{\rule{.8pt}{.8pt}}
\multiput(180,190)(.3,-.6){40}{\rule{.8pt}{.8pt}}
\multiput(180,189)(.3,-.6){23}{\rule{.9pt}{.9pt}}
\multiput(180.6,189)(.3,-.6){23}{\rule{.9pt}{.9pt}}
\multiput(180,175)(.3,0){25}{\rule{.8pt}{.8pt}}
\multiput(180,175)(-.3,0){25}{\rule{.8pt}{.8pt}}
\put(180,193){\small $\kappa_{16}$}
\put(160,175){\small $\kappa_{68}$}
\put(190,175){\small $\kappa_{78}$}
\multiput(186,183)(-.6,-.3){8}{\rule{.9pt}{.9pt}}
\multiput(186,183.6)(-.6,-.3){8}{\rule{.9pt}{.9pt}}
\put(175,155){\bf $(c)$}

\multiput(20,125)(-.3,.3){40}{\rule{.8pt}{.8pt}}
\multiput(20,125)(-.3,-.3){9}{\rule{.8pt}{.8pt}}
\multiput(20,125)(.3,0){60}{\rule{.8pt}{.8pt}}
\multiput(12,117)(-.3,-.3){15}{\rule{.8pt}{.8pt}}
\put(0,110){\small 2}
\put(0,135){\small 1}
\put(35,117){\small 3}
\put(15,120){\circle{8}}
\put(15,100){\bf $(d)$}

\multiput(100,125)(-.3,.3){40}{\rule{.8pt}{.8pt}}
\multiput(100.2,124.8)(-.3,-.3){9}{\rule{.9pt}{.9pt}}
\multiput(99.7,125.3)(-.3,-.3){9}{\rule{.9pt}{.9pt}}
\multiput(100,125)(.3,0){60}{\rule{.8pt}{.8pt}}
\multiput(92,117)(-.3,-.3){15}{\rule{.8pt}{.8pt}}
\put(95,120){\circle{8}}
\put(80,137){\small $x_1$}
\put(85,124){\small $x$}
\put(99,111){\small $1$-$x$}
\put(80,110){\small $x_2$}
\put(115,116){\small $x_3$}
\multiput(97.2,125.7)(.3,-.3){13}{\rule{.9pt}{.9pt}}
\multiput(96.8,125.3)(.3,-.3){13}{\rule{.9pt}{.9pt}}
\put(95,100){\bf $(e)$}

\multiput(180,125)(-.3,.3){40}{\rule{.8pt}{.8pt}}
\multiput(180,125)(-.3,-.3){40}{\rule{.8pt}{.8pt}}
\multiput(180,125)(.3,0){60}{\rule{.8pt}{.8pt}}
\put(180,128){\small $\kappa_{12}$}
\put(175,120){\circle*{5}}
\put(175,100){\bf $(f)$}

\multiput(20,70)(-.3,.3){40}{\rule{.8pt}{.8pt}}
\multiput(20,70)(-.3,-.3){40}{\rule{.8pt}{.8pt}}
\multiput(20,70)(.3,0){20}{\rule{.8pt}{.8pt}}
\multiput(34,70)(.3,0){15}{\rule{.8pt}{.8pt}}
\put(30,70){\circle{8}}
\put(15,45){\bf $(g)$}

\multiput(100,70)(-.3,.3){40}{\rule{.8pt}{.8pt}}
\multiput(100,70)(-.3,-.3){40}{\rule{.8pt}{.8pt}}
\multiput(100,70.3)(.3,0){20}{\rule{.9pt}{.9pt}}
\multiput(100,69.7)(.3,0){20}{\rule{.9pt}{.9pt}}
\multiput(114,70)(.3,0){15}{\rule{.8pt}{.8pt}}
\put(110,70){\circle{8}}
\multiput(102.4,72.75)(0,-.3){20}{\rule{.9pt}{.9pt}}
\multiput(103,72.75)(0,-.3){20}{\rule{.9pt}{.9pt}}
\put(95,45){\bf $(h)$}

\multiput(180,70)(-.3,.3){40}{\rule{.8pt}{.8pt}}
\multiput(180,70)(-.3,-.3){40}{\rule{.8pt}{.8pt}}
\multiput(180,70)(.3,0){60}{\rule{.8pt}{.8pt}}
\put(190,70){\circle*{5}}
\put(175,45){\bf $(i)$}

\multiput(20,20)(-.3,.3){40}{\rule{.8pt}{.8pt}}
\multiput(20,20)(-.3,-.3){40}{\rule{.8pt}{.8pt}}
\multiput(20,20)(.3,0){60}{\rule{.8pt}{.8pt}}
\put(20,20){\circle*{5}}
\put(15,-5){\bf $(j)$}

\put(0,-30){Fig. 2~~Graphical illustration of Eq. (B1).}

\end{picture}}
\end{center}
\vskip1.7in

\noindent In all terms, the external gluon quantum numbers are labeled
in the same way, 1 and 2 refer to the created gluons and 3 to the
annihilated one.  The intermediate momenta are assigned numbers 6, 7
and 8. For example, using notation from Appendix A, we have $k_1^\perp
= x_{1/7} k_7^\perp + \kappa^\perp_{16}$, where $\kappa^\perp_{16}
\equiv
\kappa^\perp_{1/7}$, $x_{1/7} \equiv x_1/x$.  Also, $\kappa^\perp_{68} =
\kappa^\perp - (1-x) \kappa^\perp_{12} /x_2$, $ \kappa^\perp_{16} = -
x_1 \kappa^\perp /x + \kappa^\perp_{12}$ and $ \kappa^\perp_{78} =
\kappa^\perp$.  All coefficients of creation operators ought to be
symmetrized with respect to quantum numbers of gluons 1 and 2, and
only one way of contracting the intermediate gluon annihilation and
creation operators is included, so that the displayed numerical
weights represent the result of all possible contractions.  Thus, in
Fig.  2.b, only one ordering of vertices from $\beta_{\infty22}$ is
included and the factor 2 is put in front.  The black dots in Fig.  2
indicate counterterms, $\beta_{\infty 11}$ in cases f and i, and
$\gamma_{\infty 21}$ in case j. Thick lines with transversal bars
denote the combined contributions of the terms $H_{A^4}$ and
$H_{[\partial A A]^2}$ from Eqs.  (A3) and (A4), $ \beta_{\infty 22} =
X_{A^4} + X_{[\partial A A]^2}$ and $
\beta_{\infty 31} = \Xi_{A^4} + \Xi_{[\partial A A]^2}$.  Since these
terms are independent of the transverse momenta and the three-gluon
vertices $\alpha_{12}$ and $\alpha_{21}$ are odd in the transverse
momentum, both terms $e$ and $h$ in Eq.  (B2), represented by diagrams
$e$ and $h$ in Fig.  2, are equal zero.

The non-zero terms are listed in the order of their appearance in Eq.
(B2), using convention $ \gamma_{21} = \sum_{n} \gamma_{21(n)}$, with
$n$ ranging from $a$ to $j$, and

$$ \gamma_{21(n)} =  \sum_{123}\int[123] \not\!\delta(k_1 + k_2 -
k_3)\,\, { g^3 \over 16 \pi^3} \,\, {1 \over 2} \,\, \gamma_{(n)}
\,\,\, a^\dagger_1 a^\dagger_2 a_3 \quad . \eqno(B3) $$

$$ \gamma_{(a)} =  8 \, {N_c \over 2} i f^{c_1 c_2 c_3}
\int_{x_1}^1{dx \,r_{\delta t}(x) \over x(1-x)(x-x_1)} \,
\int d^2\kappa^\perp \,\,r_{\Delta t}(\kappa^\perp)\,\,
{ {\cal F}_{3\lambda(a)} \over k^{+ \, 2}_3 }
\,\, \kappa^i_{68} \kappa^j_{16} \kappa^k \,\, \varepsilon^{ijk}_{(a)}
\, + \, (1 \leftrightarrow 2) \quad , \eqno(B4) $$

\noindent where

$$ r_{\delta t}(x) = r_\delta(x) \, r_\delta(1-x) \, r_\delta(x_1/x)
\, r_\delta[(x-x_1)/x] \, r_\delta[(x-x_1)/x_2] \, r_\delta[(1-x)/x_2]
\quad , \eqno(B4a) $$

$$ r_{\Delta t}(\kappa^\perp) = \exp{[-2(\kappa^{\perp \,2}_{68} +
\kappa^{\perp \,2}_{16} + \kappa^{\perp \,2})/\Delta^2]}
\quad , \eqno(B4b) $$

\begin{eqnarray*}
 & & { {\cal F}_{3\lambda(a)} \over k^{+ \, 2}_3 }  = { -x {\cal
M}^2_{16} + {\cal M}^2 \over {\cal M}^4_{16}+ {\cal M}^4 }
\left\{
\left(
{\cal M}^2_{bd} + x_2 {\cal M}^2_{68}
\right)
\right.\\
 & &
\left.
\left[
{f_{68}f_{bd}f_{16}f - 1 \over {\cal M}^4_{68}+ {\cal M}^4_{bd} +
{\cal M}^4_{16}+ {\cal M}^4 } - {f_{68}f_{bd} - 1 \over {\cal
M}^4_{68}+ {\cal M}^4_{bd}  }
\right]
\right. \\
 & &
\left.
- {{\cal M}^4_{16} + {\cal M}^4 \over {\cal M}^4_{bd}}
\left[
{f_{68}f_{16}f - 1 \over {\cal M}^4_{68}+{\cal M}^4_{16}+ {\cal M}^4 }
- {f_{68}f_{bd}f_{16}f - 1 \over {\cal M}^4_{68}+ {\cal M}^4_{bd} +
{\cal M}^4_{16}+ {\cal M}^4 }
\right]
\right\} \\
 & & + { x {\cal M}^2_{16} + x_2{\cal M}^2_{68} \over {\cal M}^4_{16}+
{\cal M}^4_{68} }
\left\{
\left(2{\cal M}^2 - {\cal M}^2_{12}
\right)
\left[
{f f_{ca}f_{16}f_{68} - 1 \over {\cal M}^4 + ({\cal M}^2 - {\cal
M}^2_{12})^2 + {\cal M}^4_{16}+ {\cal M}^4_{68} }
\right.
\right.\\
 & &
\left.
\left.
- {f f_{ca} - 1 \over {\cal M}^4+ ({\cal M}^2 - {\cal M}^2_{12})^2 }
\right]
- {{\cal M}^4_{16} + {\cal M}^4_{68} \over {\cal M}^2 - {\cal
M}^2_{12}  }
\left[
{f f_{16} f_{68} - 1 \over {\cal M}^4 + {\cal M}^4_{16}+ {\cal
M}^4_{68} }
\right.
\right. \\
 & &
\left.
\left.
- {f f_{ca}f_{16}f_{68} - 1 \over {\cal M}^4 + ({\cal M}^2 - {\cal
M}^2_{12})^2 + {\cal M}^4_{16}+ {\cal M}^4_{68} }
\right]
\right\} \quad ,
\end{eqnarray*}
$$ \eqno(B4c) $$

$$ {\cal M}^2 = {\kappa^{\perp \, 2} \over x(1-x)} \quad , \eqno(B4d)$$

$$ {\cal M}^2_{68} = { x_2^2 \kappa^{\perp \, 2}_{68} \over
(x-x_1)(1-x)} \quad , \eqno(B4e)$$

$$ {\cal M}^2_{16} = { x^2 \kappa^{\perp \, 2}_{16} \over x_1(x-x_1)}
\quad , \eqno(B4f)$$
\vskip.1in

\noindent with $f_u = \exp{[-u^2/\lambda^4]}$, $bd = {\cal M}^2_{bd} =
{\cal M}^2_{68}/x_2 + {\cal M}^2_{12}$, $ca = {\cal M}^2 - {\cal
M}^2_{12}$, $f \equiv f_{cd}$, $cd = {\cal M}^2$, and

\begin{eqnarray*}
 & &\varepsilon^{ijk}_{(a)} = \varepsilon_1^{*j} \varepsilon_2^{*i}
\varepsilon_3^k
\left[ 1 - {x\over x - x_1} + {1 \over x_1} - {2 x \over x_1} +
{ x \over (1 - x) x_1} + {x x_2 \over (1 - x) x_1} + {x x_2 \over (x -
x_1) x_1} \right] \\ & & + \varepsilon_1^{*k} \varepsilon_2^{*i}
\varepsilon_3^j
\left[ {1 \over x - x_1} - {1 \over 1 - x} \right]
+ \varepsilon_1^{*k} \varepsilon_2^{*j} \varepsilon_3^i
\left[ {- x_2\over (1 - x) (x - x_1)} \right]
+ \varepsilon_1^{*i} \varepsilon_2^{*k} \varepsilon_3^j
\left[ {x_2 \over (1 - x) (x - x_1)} \right] \\
 & & + \varepsilon_1^{*i} \varepsilon_2^{*j} \varepsilon_3^k
\left[{-x_2 \over x - x_1} + {x x_2 \over (1 - x) (x - x_1)} \right]
+ \varepsilon_1^{*j} \varepsilon_2^{*k} \varepsilon_3^i
\left[{-x_2 \over (1 - x) x_1} - {x x_2 \over (1 - x) (x - x_1) x_1} \right]\\
 & & + \varepsilon_1^* \varepsilon_2^*
\left[\delta^{ik} \varepsilon_3^j {x_2 \over (1 - x)^2} +
\delta^{jk} \varepsilon_3^i {x_2 \over (1 - x) x} +
\delta^{ij} \varepsilon_3^k \left({x x_2 \over (x - x_1)^2} -
 {x_2 \over 1 - x}\right) \right] \\ & & + \varepsilon_1^*\varepsilon_3
\left[ \delta^{jk} \varepsilon_2^{*i} \left({x \over (1 - x) (x - x_1)}
- { 1 \over x}\right) -
\delta^{ij} \varepsilon_2^{*k} {x x_2 \over (1 - x) (x - x_1)^2} -
\delta^{ik} \varepsilon_2^{*j} {x x_2 \over (1 - x)^2 (x - x_1)} \right] \\
 & & +  \varepsilon_2^*\varepsilon_3
\left[\delta^{ij} \varepsilon_1^{*k} {- x_2 \over (x - x_1)^2} +
\delta^{jk} \varepsilon_1^{*i} {x_2 \over x (x - x_1)} -
\delta^{ik} \varepsilon_1^{*j}\left( {x x_2 \over (1 - x)^2 x_1} -
{x_2 \over (x - x_1) x_1}\right) \right]
\quad .
\end{eqnarray*}
$$ \eqno(B4g) $$

$$ \gamma_{(b)} =  2 \, {N_c\over 2} i f^{c_1 c_2 c_3}
\int_{x_1}^1{dx\,r_{\delta t}(x) \over x(1-x)} \,
\int d^2\kappa^\perp \,\,r_{\Delta t}(\kappa^\perp) \,\,
{ {\cal F}_{2\lambda(b)} \over k^+_3 }\,\,\varepsilon_{(b)} \, + \, (1
 \leftrightarrow 2) \quad  \eqno(B5) $$

\noindent where

$$ { {\cal F}_{2\lambda(b)} \over k^{+ \, 2}_3 } \,  =  \, { 2{\cal
M}^2 - {\cal M}^2_{12} \over ({\cal M}^2 - {\cal M}^2_{12})^2 + {\cal
M}^4 } \, ( f_{ac}f - 1 ) \quad , \eqno(B5a) $$

\begin{eqnarray*}
 & & \varepsilon_{(b)} \equiv  \varepsilon^\perp_{(b)}\kappa^\perp =
 \\ & &
\varepsilon_1^*\varepsilon_2^* \, \varepsilon_3\kappa
\left( 1 - s_{(b)} - {1 \over x} - {1 \over 1-x} \right)
+
\varepsilon_1^*\varepsilon_3 \, \varepsilon_2^*\kappa
\left({1\over x} + {s_{(b)} \over 1-x}\right)
+
\varepsilon_2^*\varepsilon_3 \, \varepsilon_1^*\kappa
\left({1 \over 1 - x } + {s_{(b)} \over x}\right)
\, ,
\end{eqnarray*}
$$ \eqno(B5b) $$

\noindent and $ s_{(b)} = (x_1 + x)(x_2 + 1 -x)/(x - x_1)^2$.

$$ \gamma_{(c)} =  2 \, { - N_c\over 2} i f^{c_1 c_2 c_3}
\int_{x_1}^1{dx\,r_{\delta t}(x) \over (x-x_1)(1-x)} \,
\int d^2\kappa^\perp \,\,r_{\Delta t}(\kappa^\perp) \,\,
{ {\cal F}_{2\lambda(b)} \over k^+_3 }\,\,\varepsilon_{(c)} \, + \, (1
 \leftrightarrow 2) \quad  \eqno(B6) $$

\noindent where

$$ { {\cal F}_{2\lambda(c)} \over k^+_3 } \,  =  \, { x_2{\cal
M}^2_{68} + {\cal M}^2_{bd} \over {\cal M}^4_{68} + {\cal M}^4_{bd}}
\, ( f_{68}f_{bd} - 1 ) \quad , \eqno(B6a) $$

\begin{eqnarray*}
 & &\varepsilon_{(c)} \equiv \varepsilon^\perp_{(c)}\kappa^\perp_{68} =
\varepsilon_1^*\varepsilon_2^* \, \varepsilon_3\kappa_{68}
\left( {-x_2 \over x-x_1} +{ x_2 s_{(c)} \over 1-x} \right)\\
 & & +
\varepsilon_1^*\varepsilon_3 \, \varepsilon_2^*\kappa_{68}
\left(- 1 - s_{(c)} + {x_2 \over 1-x} + {x_2 \over x-x_1}\right)
+
\varepsilon_2^*\varepsilon_3 \, \varepsilon_1^*\kappa_{68}
\left({-x_2 \over 1 - x } + {x_2 s_{(c)} \over x-x_1}\right)
\, ,
\end{eqnarray*}
$$ \eqno(B6b) $$

\noindent and $ s_{(c)} = (x_1 - x + x_1)(1 -x + 1)/x^2$.

\begin{eqnarray*}
 & &  \gamma_{(d)}+ \gamma_{(f)} =  \\ & & 4 \, N_c Y_{123}
\int_0^1{dx\,r_{\delta \mu}(x) \over x(1-x)} \,
\int d^2\kappa^\perp \,\,r_{\Delta \mu}(\kappa^\perp) \,\,
\kappa^{\perp \, 2}\left[1 + {1\over x^2} + {1\over (1-x)^2}\right]
\left[
{ {\cal F}_{3\lambda(d)} \over x^2_2 k^{+\, 2}_3 } + { {\cal
F}_{2\lambda(f)} \over  x_2 k^+_3 {\cal M}^2 }
\right] \\
 & & +4 \, Y_{123}\, { {\cal F}_{2\lambda(f)} \over  x_2 k^+_3
}{\tilde \mu}^2_\delta \, + \, (1 \leftrightarrow 2) \quad ,
\end{eqnarray*}
$$ \eqno(B7) $$

\noindent where

\begin{eqnarray*}
 & & { {\cal F}_{3\lambda(d)} \over x^2_2 k^{+\, 2}_3 }  + { {\cal
F}_{2\lambda(f)} \over  x_2 k^+_3 {\cal M}^2 } = {1\over x_2^2} {
{\cal M}^2_{12} - x_2 {\cal M}^2 \over {\cal M}^4_{12}+ {\cal M}^4 }
\left\{
\left[
{\cal M}^2 \left({1\over x_2}+x_2 \right) + {\cal M}^2_{12}\right]
\right.\\
 & &
\left.
\left[
{f^2 f_{bd}f_{12} - 1 \over 2{\cal M}^4 + ({\cal M}^2/ x_2 +{\cal
M}^2_{12})^2 + {\cal M}^4_{12} } - {f f_{bd} - 1 \over {\cal M}^4 +
({\cal M}^2/ x_2 +{\cal M}^2_{12})^2 }
\right]
\right. \\
 & &
\left.
- {{\cal M}^4 + {\cal M}^4_{12} \over {\cal M}^2/ x_2 + {\cal
M}^2_{12}}
\left[
{f^2 f_{12} - 1 \over 2{\cal M}^4 +{\cal M}^4_{12} } - {f^2
f_{bd}f_{12} - 1 \over  2{\cal M}^4 + ({\cal M}^2/ x_2 +{\cal
M}^2_{12})^2 + {\cal M}^4_{12} }
\right]
\right\} \\
 & & + { {\cal M}^2_{12}  \over x_2 {\cal M}^2 } {f_{12}f^2  - 1 \over
{\cal M}^4_{12} + 2 {\cal M}^4}
\quad ,
\end{eqnarray*}
$$ \eqno(B7a) $$

\noindent ${\cal M}_{bd}^2 = {\cal M}^2/ x_2 +{\cal M}^2_{12}$, since
${\cal M}_{68}^2 \equiv {\cal M}^2$ in terms $d$ and $f$ (here $f$ is
a subscript, not a form factor), $f_{12} \equiv f_{ad}$, $ad = {\cal
M}^2_{12}$,

$$ { {\cal F}_{2\lambda(f)} \over x_2 k^+_3 } \,  =  \, { f_{12} - 1
\over x_2 {\cal M}^2_{12}} \quad , \eqno(B7b) $$

\noindent $r_{\delta \mu}(x)$ is given in Eq.  (4.7), $ r_{\Delta
\mu}(\kappa^\perp) = r^4_\Delta (\kappa^{\perp \, 2})$, and $ 2g^2
{\tilde \mu}^2_\delta = 16\pi^3 \mu^2_\delta$.

\begin{eqnarray*}
 & & \gamma_{(g)} + \gamma_{(i)}  =  \\ & & 2 \, N_c Y_{123}
\int_0^1{dx\,r_{\delta \mu}(x) \over x(1-x)} \,
\int d^2\kappa^\perp \,\,r_{\Delta \mu}(\kappa^\perp) \,\,
\kappa^{\perp \, 2}\left[1 + {1\over x^2} + {1\over (1-x)^2}\right]
\left[
{ {\cal F}_{3\lambda(g)} \over k^{+\, 2}_3 } + { {\cal
F}_{2\lambda(i)} \over k^+_3 {\cal M}^2 }
\right] \\
 & & +2 \, Y_{123}\, { {\cal F}_{2\lambda(i)} \over k^+_3 }{\tilde
\mu}^2_\delta \, + \, (1 \leftrightarrow 2) \quad ,
\end{eqnarray*}
$$ \eqno(B8) $$

\noindent where

\begin{eqnarray*}
 & & { {\cal F}_{3\lambda(g)} \over k^{+\, 2}_3 }  + { {\cal
F}_{2\lambda(i)} \over k^+_3 {\cal M}^2 } = { - {\cal M}^2_{12}\over
{\cal M}^2 } {f_{12} f^2  - 1 \over {\cal M}^4_{12} + 2{\cal M}^4 } -{
{\cal M}^2 + {\cal M}^2_{12}  \over {\cal M}^4 + {\cal M}^4_{12} }
\left\{
\left( 2{\cal M}^2 - {\cal M}^2_{12}\right)
\right.\\
 & &
\left.
\left[
{f^2 f_{ca}f_{12} - 1 \over 2{\cal M}^4 + ({\cal M}^2 - {\cal
M}^2_{12})^2 + {\cal M}^4_{12} } - {f f_{ca} - 1 \over {\cal M}^4 +
({\cal M}^2 - {\cal M}^2_{12})^2 }
\right]
\right. \\
 & &
\left.
+ {{\cal M}^4 + {\cal M}^4_{12} \over {\cal M}^2_{12} - {\cal M}^2 }
\left[
{f^2 f_{12} - 1 \over 2{\cal M}^4 +{\cal M}^4_{12} } - {f^2
f_{ca}f_{12} - 1 \over  2{\cal M}^4 + ({\cal M}^2 - {\cal M}^2_{12})^2
+ {\cal M}^4_{12} }
\right]
\right\}
\quad ,
\end{eqnarray*}
$$ \eqno(B8a) $$

\noindent with $ca = {\cal M}^2 - {\cal M}^2_{12}$, and

$$ { {\cal F}_{2\lambda(i)} \over k^+_3 } \,  =  \, - { f_{12} - 1
\over {\cal M}^2_{12}} \quad . \eqno(B8b) $$

$ \gamma_{(j)} = \gamma_\infty + (1 \leftrightarrow 2)$, where
$\gamma_\infty$ denotes the counterterm coefficient in

$$ \gamma_{\infty 21} =
\sum_{123}\int[123] \not\!\delta(k_1 + k_2 - k_3)\,\,
{ g^3 \over 16 \pi^3} \,\, \gamma_\infty
\,\,\, a^\dagger_1 a^\dagger_2 a_3 \quad , \eqno(B9) $$

\noindent whose dependence on the quantum numbers $1$, $2$ and $3$ needs
to be found.

\vskip.3in
{\bf Appendix C: Ultraviolet divergent terms in $\gamma_{\lambda 21}$}
\vskip.1in

The three-gluon vertex is given by Eqs.  (B4-9).  The ultraviolet
diverging parts of terms $b$ and $c$ vanish.  This can be seen by
changing variables so that the sum of squares of relative transverse
momenta in three subsequent vertices in Eq.  (B4b), is written as

$$ \eta_{68} \kappa^{\perp \,2}_{68} + \eta_{16} \kappa^{\perp
\,2}_{16} + \eta \kappa^{\perp \,2} =  \zeta \rho^{\perp \, 2} + \chi
\kappa^{\perp
\, 2}_{12} \quad , \eqno(C1) $$

\noindent where $ \zeta = \eta_{68} + \eta_{16} x_1^2/x^2 + \eta$, $
\rho^\perp = \kappa^\perp - \xi \kappa^\perp_{12}$, $ \xi = [\eta_{68}
(1-x)/x_2 + \eta_{16} x_1/x ]/ \zeta$, and $ \chi = [
\eta_{68}\eta_{16} (x-x_1)^2/(x x_2)^2 + \eta_{68} \eta (1-x)^2/x_2^2
+ \eta_{16} \eta ]/
\zeta$.  The coefficients $\eta$ are introduced for identification of
finite parts of counterterms that may contain functions of $x_1$.  All
$\eta$s are equal 2 in Eq.  (B4b).  Using variable $\rho^\perp$, the
potentially diverging relative transverse momentum integrals in terms
$b$ and $c$, can be written as

$$ \gamma_{(b) div} = - 2 \, {N_c\over 2} i f^{c_1 c_2 c_3}
\int_{x_1}^1 dx\,r_{\delta t}(x) \, \int d^2\rho^\perp
\,\,e^{-\zeta \rho^{\perp \, 2}/\Delta^2} \,\,
{\kappa^\perp \over \kappa^{\perp \, 2} }\,\varepsilon^\perp_{(b)} \,
 + \, (1 \leftrightarrow 2) \quad ,
\eqno(C2a) $$

$$ \gamma_{(c) div} = 2 \, { N_c\over 2} i f^{c_1 c_2 c_3}
\int_{x_1}^1 dx\,r_{\delta t}(x) \, \int d^2\rho^\perp \,\,e^{-\zeta
\rho^{\perp \, 2}/\Delta^2} \,\, { \kappa^\perp_{68} \over
\kappa^{\perp \, 2}_{68}  \, x_2 } \,\varepsilon^\perp_{(c)} \, + \, (1
\leftrightarrow 2) \quad . \eqno(C2b) $$

\noindent Both terms contain an integral of the same structure,

$$ I^\perp = \int d^2 \kappa^\perp \, {\kappa^\perp \over \kappa^2} \,
\exp{[-\zeta (\kappa^\perp - v^\perp)^2/\Delta^2]} \quad , \eqno(C2c)
$$

\noindent and $ \lim_{\Delta \rightarrow \infty} I^\perp = \pi v^\perp$,
which is a finite ultraviolet regularization effect with no
divergence.  Thanks to the regularization factors $r_{\delta t}$,
these terms can be integrated over $x$ and the resulting function of
$x_1$ depends on the choice of the coefficients $\eta$. If they were
chosen to depend on $x$, an arbitrarily complex, ultraviolet
regularization dependent finite function of $x_1$ can be obtained, but
$\gamma_{(b) div} = \gamma_{(c) div} = 0$.

\begin{eqnarray*}
 & & \gamma_{(g+i)div}  =  \\ & & - N_c Y_{123}
\int_0^1 dx\,r_{\delta \mu}(x)
\int_{\mu^2}^\infty { \pi d\kappa^2 \over \kappa^2 }
\,\,e^{-4\kappa^2/\Delta^2} \,\,
x(1-x)\left[1 + {1\over x^2} + {1\over (1-x)^2}\right] \, + \, (1
 \leftrightarrow 2) \quad .
\end{eqnarray*}
$$ \eqno(C3) $$

\noindent $\mu^2$ is an arbitrary parameter inserted here to simplify
notation for the well defined integral over $\kappa^2$ at small
$\kappa$ when the terms with form factors are written explicitly and
the small denominator effect is absent.  One also obtains

$$ \gamma_{(d+f)div} = 2 \gamma_{(g+i)div} \quad . \eqno(C4) $$

\noindent The logarithmically diverging term (C3) appears then with
factor $-3$ in the counterterm coefficient $\gamma_\infty$ in Eq.
(B9).  Its finite contribution to the counterterm is a finite number
times $Y_{123}$.

The remaining term, $\gamma_{(a)}$ of Eq.  (B4), contains the tensor
$\kappa^{ijk} = \kappa^i_{68} \kappa^j_{16} \kappa^k$ contracted with
$\varepsilon^{ijk}_{(a)}$ and the latter does not depend on transverse
momenta.  The ultraviolet regularization dependence comes only from

$$ \left[{ {\cal F}_{3\lambda(a)} \over k^{+ \, 2}_3 }\right]_{\Delta}
= {x_2 \over {\cal M}^2 {\cal M}^2_{68}} \quad . \eqno(C5) $$

\noindent and using Eq.  (C1) one obtains

$$ \gamma_{(a) div} =  8 \, {N_c \over 2} i f^{c_1 c_2 c_3}
\int_{x_1}^1 dx \,r_{\delta t}(x) {1-x \over x_2} \,
I^{ijk} \varepsilon^{ijk}_{(a)}
\, + \, (1 \leftrightarrow 2) \quad , \eqno(C6) $$

\noindent where

$$ I^{ijk} = \int d^2\rho^\perp \,\,e^{-\zeta \rho^{\perp \,
2}/\Delta^2}\,\, { \kappa^i_{68} \kappa^j_{16} \kappa^k  \over
\kappa^{\perp \, 2}
\kappa^{\perp \, 2}_{68}}
 \quad , \eqno(C7) $$

\noindent The ultraviolet-regularization-dependent part of $I^{ijk}$
can be evaluated using Feynman's trick to combine denominators and
making similar steps as in Eq. (C2),

$$ \left[I^{ijk}\right]_\Delta = {\pi \over 2} {x_1 \over x} {1-x
\over x_2} \left[\kappa_{12}^i \delta^{jk} -
\kappa_{12}^k \delta^{ij} + { x-x_1 + x x_2 \over x_1(1-x)}
\kappa_{12}^j \delta^{ik} \right] \ln{\Delta \over |\kappa^\perp_{12}|}
\quad . \eqno(C8) $$

\noindent A number of finite terms are obtained that depend on the
choice of coefficients $\eta$ in Eq.  (C1).  The terms include also a
new tensor structure
$\kappa_{12}^i\kappa_{12}^j\kappa_{12}^k/\kappa_{12}^2$.  Inserting
$\left[I^{ijk}\right]_\Delta$ in place of $I^{ijk}$ in Eq. (C6), one
obtains

$$ \gamma_{(a) div} =  8 \, {N_c \over 2}i f^{c_1 c_2 c_3}
\int_{x_1}^1 dx \,r_{\delta t}(x) {1-x \over x_2} \,
\left[I^{ijk}\right]_\Delta \varepsilon^{ijk}_{(a)}
\, + \, (1 \leftrightarrow 2) \quad , \eqno(C9) $$

\noindent where

$$ i f^{c_1 c_2 c_3}{1-x \over x_2}
\left[I^{ijk}\right]_\Delta \varepsilon^{ijk}_{(a)} =
\pi \ln{\Delta \over |\kappa^\perp_{12}|}  \left[ c_{12}
Y_{12} + V_{13} Y_{13} + c_{23} Y_{23} \right] \quad , \eqno(C9a) $$

\noindent and

$$ c_{12} = {2 \over 1-x} + {1 \over x-x_1} + {1 \over x} + {(1-x)^2
\over x_2^2 } - {2 \over x_2} \quad , \eqno(C9b) $$

$$ c_{13} = {2 \over 1-x} + {1 \over x-x_1} + {1 \over x} + {(1-x)^2
\over x_2 } - 2 \quad , \eqno(C9c) $$

$$ c_{23} = {2 \over 1-x} + {1 \over x-x_1} + {1 \over x} - {(1-x)^2
\over x_2^2 } - { 1 + x^2 \over x_2} - 2 \quad , \eqno(C9d) $$

\noindent with

$$ Y_{12} = i f^{c_1 c_2 c_3}\varepsilon_1^*\varepsilon_2^* \cdot
\varepsilon_3 \kappa_{12} \quad , \eqno(C10a) $$

$$ Y_{13} = - i f^{c_1 c_2 c_3}\varepsilon_1^*\varepsilon_3 \cdot
\varepsilon_2^* \kappa_{12} {1\over x_{2/3}} \quad , \eqno(C10b) $$

$$ Y_{23} = - i f^{c_1 c_2 c_3}\varepsilon_2^*\varepsilon_3 \cdot
\varepsilon_1^* \kappa_{12} {1\over x_{1/3}} \quad . \eqno(C10c) $$

\noindent The counterterm should cancel the divergence, so that

$$ 0 = \gamma_{(a) div} + \gamma_{(d+f)div} + \gamma_{(g+i)div} +
\gamma_{\infty div} + (1 \rightarrow 2) $$

$$ = 2 N_c \pi \ln{ \Delta \over \mu}
\left\{
2 \int_{x_1}^1 dx \,r_{\delta t}(x)
\left[ c_{12} Y_{12} + c_{13} Y_{13} + c_{23} Y_{23} \right]
\right. $$
$$ \left.-3 \left[ Y_{12} + Y_{13} + Y_{23} \right]
\int_0^1 dx\,r_{\delta \mu}(x)
x(1-x)\left[1 + {1\over x^2} + {1\over (1-x)^2}\right]\right\} +
\gamma_{\infty div} + (1 \leftrightarrow 2)  \quad , \eqno(C11)$$

\noindent and

$$  \gamma_{(j) div} = - Y_{123} {2 N_c \pi \over 3} \ln{ \Delta \over
\mu}
\left[ 11 + h(x_1) \right]  \quad , \eqno(C12) $$

\noindent where

$$ h(x_1) = 6 \int_{x_1}^1 dx \,r_{\delta t}(x) \left[ {2 \over 1-x} +
{1 \over x-x_1} + {1 \over x} \right] - 9 \int_0^1 dx \,r_{\delta
\mu}(x) \left[ {1\over x} + {1 \over 1-x} \right] + (1 \leftrightarrow
2) \quad . \eqno(C13) $$

The ultraviolet counterterm $\gamma_{\infty 21}$ has the form (B9),
where

$$ \gamma_\infty = Y_{123} { - N_c \pi \over 3} \, \ln{ \Delta \over
\mu} \left[ 11 + h(x_1) \right]  + \gamma_{finite}
\quad . \eqno(C14) $$

\noindent Different choices of the coefficients $\eta$ in Eq.  (C1),
lead to different finite terms $\gamma_{finite}$, which demonstrates
finite dependence on ultraviolet regularization. To restore covariance
of observables, the finite terms must be then allowed to contain
unknown functions of $x_1$ that multiply three structures $Y_{12}$,
$Y_{13}$, and $Y_{23}$ from Eq.  (C10), and the fourth structure,

$$ Y_4 = i f^{c_1 c_2 c_3} \varepsilon^*_1 \kappa_{12}
\cdot \varepsilon^*_2 \kappa_{12} \cdot \varepsilon_3 \kappa_{12} /
\kappa_{12}^2 \quad . \eqno(C15) $$

Inclusion of $n_f$ flavors of quarks produces in $\gamma_{(j)div}$
additional diverging terms of the form

$$ - 2\pi n_f \ln{\Delta \over \mu} \left[ c_{f12} Y_{12} + c_{f13}
Y_{13} + c_{f23} Y_{23} + c_{fm} Y_{123} \right]
\quad , \eqno(C16) $$

\noindent where

$$  c_{f12} = - \int_{x_1}^1 dx { 1 - 4x + 2 x^2 + x_1 \over 2 x_2^2 }
+ (1 \leftrightarrow 2)
\quad , \eqno(C17a) $$

$$  c_{f13} = - \int_{x_1}^1dx {1 - 4x + 2 x^2 + x_1 \over 2 x_2} + (1
\leftrightarrow 2)
\quad , \eqno(C17b) $$

$$ c_{f23} = - \int_{x_1}^1dx { -2 + 4x - 2x^2(1+x_2) + x_1 x_2 \over
2 x_2^2} + (1 \leftrightarrow 2)
\quad , \eqno(C17c) $$

$$ c_{fm} = \int_0^1dx {-3 \over 8}2[1-2x(1-x)] + (1 \leftrightarrow 2)
\quad . \eqno(C17d) $$

\noindent After simplifications, integration and symmterization,
fermions' contribution turns out to be not sensitive to small-$x$
regularization and changes Eq.  (C14) to

$$ \gamma_{(j) div} = - Y_{123} {2 \pi \over 3} \ln{ \Delta \over \mu}
\left\{ N_c \left[11 + h(x_1) \right] - 2 n_f \right\} \quad .
\eqno(C18) $$

\vskip.3in
{\bf Appendix D: $W_\lambda(x)$ in Eq. (5.2)}
\vskip.1in

$W_{\lambda12}(x_1, \kappa_{12}^{\perp \, 2})$, $W_{\lambda13}(x_1,
\kappa_{12}^ {\perp \, 2} )$ and $W_{\lambda23}(x_1, \kappa_{12}^{\perp
\, 2})$ in Eq.  (4.14) become all equal to $W_\lambda(x_1)$ when
$\kappa_{12}^\perp = 0$.  Calculation of $W_\lambda(x_1)$ is based on
extraction of coefficients of terms linear in $\kappa^\perp_{12}$,
which form $Y_{ij}$ for $ij = 12,13,23$, and employs the following
facts.  The self-interaction terms $d$ and $g$, and mass counterterms
$f$ and $i$ from Fig.  2, contribute through

$$ {g^3 \over 16 \pi^3 } N_c \int_0^1 dx \int {d^2 \kappa^\perp \over
\kappa^{\perp \, 2} } x(1-x)\left[1 + {1 \over x^2} + {1 \over
(1-x)^2}\right] \times $$

$$ \left[ 5 f^2 + 4 f f_{bd} - 4f^2 f_{bd} - 2 f^3 - 3\theta(\mu^2 -
\kappa^2) \right] \quad . \eqno(D1) $$

\noindent $\mu$ is introduced only for simplification and is canceled in
the full formula for dependence of $W_\lambda(x_1)$ on $\lambda$.  In
the term $\gamma_{(a)}$, the renormalization group factor of Eq.
(B4c) contributes to the dependence of $W_\lambda(x_1)$ on $\lambda$
only through small $A$ expansion of

$$ \left[{ {\cal F}_{3\lambda(a)} \over k^{+ \, 2}_3 }\right]_\lambda
= - {x_2 \over [{\cal M}^2_{680} - 2 x_2 A] {\cal M}^2 } \, f f_{16}
f_{68} \quad , \eqno(D2) $$

\noindent where $A = \kappa^\perp \kappa^\perp_{12}/(x-x_1)$ and the
added subscript $0$ indicates that $\kappa^\perp_{12}$ is set equal 0.
Other parts do not contribute because when the form factors $f$ are
expanded, no dependence on $\lambda$ is generated due to dimensional
reasons and the remaining factors cannot contribute since they are
multiplied by differences of form factors and there is an identity

$$ \int_0^\infty {dz\over z} \left[ e^{-az^2/\lambda^4} -
e^{-bz^2/\lambda^4} \right] = {1\over 2} \ln{b \over a} , \eqno(D3) $$

\noindent which shows that $\lambda$ drops out.  The expansion in $A$ in
Eq.  (D2), produces the same tensor structure as in Eq.  (C8), which
leads then to Eq.  (5.2).

\vskip.3in
{\bf Appendix E: Regularization mixing for $x$ and $\kappa^\perp$ }
\vskip.1in

Regularization with invariant masses implies that the coefficients
$\eta$ in Eq.  (C1) become equal $ \eta = 1/x + 1/(1-x)$, $ \eta_{68}
= x_2/( x - x_1) + x_2/( 1-x )$ and $\eta_{16} = x/x_1 + x/(x-x_1)$.
These coefficients diverge when $x \rightarrow x_1$ or $x \rightarrow
1$.  In mass counterterms, the coefficient $\eta$ is the same.  When
integrating over transverse momenta in mass counterterms with
$r_\Delta(2\eta \kappa^{\perp \, 2})$, one obtains in Eq.  (4.9) only
logarithmically divergent integral over $x$ and a regularization
factor of the form $r_\delta(x) = x^\delta$ is sufficient to regulate
it.  In other words, in place of the second scale, $\epsilon$, in
$r_\delta$ from Eq.  (5.9), one can consider regularizations of
transverse divergences that provide additional damping of the
small-$x$ region.  The question is what are the consequences of the
mixing of large $\kappa$  and small $x$ in $\gamma_{\lambda 21}$.
Most representative is Eq.  (B4).  Although all three $\eta$s can grow
to $\infty$, $\chi$ is limited and does not exceed $1/(x_1 x_2)$
reaching this value at the ends of the integration range, while $\xi =
1$ when $x=x_1$ and drops down to 0 at $x=1$.  Therefore, $\chi
\kappa^{\perp \, 2}_{12}/\Delta^2$ in the exponent is always a small
number and vanishes when $\Delta \rightarrow
\infty$ without contributing to regularization dependence.  On the other
hand, the coefficient $\zeta$ grows to infinity at the ends of the
integration region in $x$.  This way the ultraviolet regularization
factor depending on invariant masses changes small-$x$ singularities.
The same phenomenon occurs in a simpler form in the instantaneous
terms involving $1/\partial^{+ \, 2}$.  The instantaneous terms
themselves do not contain integration over $\kappa^\perp$, but the
integration occurs when these terms are included in the dynamics.  The
logarithmically divergent integrals over $\kappa^\perp$ produce
logarithms of $\zeta$ as $\Delta$-independent remnants of the
ultraviolet regularization and one cannot exclude arbitrary functions
of $x$ in finite parts of the ultraviolet counterterms, including
integrals that strengthen small-$x$ logarithmic divergences.
Invariant mass regularizations including a small gluon mass
$\mu$~\cite{GlazekAPP12}, provide additional damping for $x \sim
\mu^2/\Delta^2$.

\end{document}